\documentclass[manuscript, acmsmall, nonacm, screen]{acmart}
\setcopyright{none}

\AtBeginDocument{%
  }

\usepackage{comment}
\usepackage{graphics}
\usepackage{color}
\usepackage{fontawesome}
\usepackage{todonotes}
\usepackage{xspace}
\usepackage{siunitx}
\usepackage[flushleft]{threeparttable}
\usepackage{enumitem}
\usepackage{balance}
\usepackage{caption}
\usepackage{graphics}
\usepackage{color, colortbl, soul, xcolor}
\usepackage{tikz}
\usepackage{caption,subcaption}
\usepackage{fixltx2e}
\usepackage{bm}
\usepackage{romannum}
\usepackage{mathtools}
\usepackage{makecell}
\usepackage{tabularx,booktabs,caption,ragged2e}
\usepackage{tikzsymbols}
\usepackage{amsmath}
\usepackage{algorithmic}
\usepackage[title]{appendix}
\usepackage{color, colortbl, soul, xcolor}
\usepackage{bm}
\usepackage{algorithm}
\usepackage{algorithmic}
\usepackage[utf8]{inputenc}
\usepackage{tcolorbox}
\usepackage{hyperref}
\usepackage{changepage}
\usepackage[normalem]{ulem}

\definecolor{subheadergray}{gray}{0.55}

\begin{document}

\title{LLM-as-a-Judge in Healthcare: A Scoping Analysis of Applications, Methods, and Human Alignment}

\author{Lingyao Li}
\orcid{0000-0001-5888-8311}
\authornote{Both authors contributed equally to this research.}
\authornote{Corresponding authors.}
\email{lingyaol@usf.edu}
\author{Deyi Li}
\orcid{0000-0003-3660-4175}
\authornotemark[1]
\authornotemark[2]
\email{lideyi@ufl.edu}
\affiliation{%
  \institution{University of South Florida}
  \city{Tampa}
  \state{FL}
  \institution{University of Florida}
  \city{Gainesville}
  \state{FL}
  \country{USA}
}

\author{Chen Chen}
\orcid{0000-0001-7179-0861}
\email{chechen@fiu.edu}
\affiliation{%
  \institution{Florida International University}
  \city{Miami}
  \state{FL}
  \country{USA}
}

\author{Renkai Ma}
\orcid{0000-0002-4434-2235}
\email{renkai.ma@uc.edu}
\affiliation{%
  \institution{University of Cincinnati}
  \city{Cincinnati}
  \state{OH}
  \country{USA}
}

\author{Runlong Yu}
\email{ryu5@ua.edu}
\orcid{0000-0003-4080-2377}
\affiliation{%
  \institution{University of Alabama}
  \city{Tuscaloosa}
  \state{AL}
  \country{USA}
}

\author{Mingquan Lin}
\email{lin01231@umn.edu}
\orcid{0009-0003-6619-7889}
\affiliation{%
  \institution{University of Minnesota}
  \city{Minneapolis}
  \state{MN}
  \country{USA}
}

\author{Rui Yin}
\email{ruiyin@ufl.edu}
\orcid{0000-0002-1403-0396}
\affiliation{%
  \institution{University of Florida}
  \city{Gainesville}
  \state{FL}
  \country{USA}
}

\author{Lizhou Fan}
\email{lizhouf@umich.edu}
\orcid{0000-0002-7962-9113}
\affiliation{%
  \institution{University of Michigan}
  \city{Ann Arbor}
  \state{MI}
  \country{USA}
}

\author{Cathy Shyr}
\email{cathy.shyr@vumc.org}
\orcid{0000-0001-7466-0034}
\affiliation{%
  \institution{Vanderbilt University Medical Center}
  \city{Nashville}
  \state{TN}
  \country{USA}
}

\author{Siyuan Ma}
\email{siyuan.ma@vumc.org}
\orcid{0000-0001-6659-3441}
\affiliation{%
  \institution{Vanderbilt University Medical Center}
  \city{Nashville}
  \state{TN}
  \country{USA}
}

\author{Mei Liu}
\authornotemark[2]
\email{mei.liu@ufl.edu}
\orcid{0000-0002-8036-2110}
\affiliation{%
  \institution{University of Florida}
  \city{Gainesville}
  \state{FL}
  \country{USA}
}

\author{Steven Bethard}
\authornotemark[2]
\email{bethard@email.arizona.edu}
\orcid{0000-0001-9560-6491}
\affiliation{%
  \institution{University of Arizona}
  \city{Tucson}
  \state{AZ}
  \country{USA}
}

\renewcommand{\shortauthors}{Li et al.}

\begin{abstract}
    \textbf{Abstract:} Large language models (LLMs) are increasingly deployed across healthcare applications, including clinical documentation, diagnostic reasoning, medicine recommendation, and medical education. Their outputs are largely unstructured clinical text, which is difficult to reliably evaluate at scale. LLM-as-a-Judge, in which an LLM evaluates another system's output against task-specific criteria, offers a scalable alternative and is increasingly used in clinical evaluation, yet its validity in healthcare remains underexamined. Existing reviews focus on general-purpose LLM evaluation or on risk framework, rather than systematically characterizing how LLM-as-a-Judge is applied in healthcare and how well their judgments align with human experts. We therefore conduct a PRISMA-guided comprehensive review of LLM-as-a-Judge applications in healthcare, searching five databases for studies published between January~2023 and February~2026. After screening 541 records, 134 studies meet the eligibility and are coded by health scenario, judge configuration, technical approach, and validation design. LLM-as-a-Judge is concentrated in clinical decision support, clinical natural language processing (NLP), medical knowledge and question answering (QA), and medical communication. OpenAI models are the most frequently used judges, and prompt engineering appears in nearly all studies, with ensemble, multi-agent, and retrieval-augmented designs as common extensions. Among studies reporting human validation, LLM judges often show moderate to strong alignment with expert judgments, although reliability varies substantially across tasks. Overall, this review positions LLM-as-a-Judge as a promising framework for scalable healthcare AI evaluation, while emphasizing that its clinical value depends on model design and rigorous validation.
\end{abstract}




\maketitle

\pagenumbering{arabic}

\section{INTRODUCTION}
\label{sec:intro}

Artificial intelligence (AI) systems, particularly large language models (LLMs), are increasingly incorporated into healthcare applications and clinical workflows, including documentation~\cite{kweon2024ehrnoteqa}, diagnostic reasoning~\cite{goh2024large}, medicine recommendation~\cite{zhou2025collaborative}, emergency service~\cite{li2026dispatchmas}, and medical education~\cite{yu2025simulated}. As these systems generate clinical text that may shape interpretation, communication, or decisions, rigorous evaluation is essential for safe deployment. However, scalable evaluation remains challenging, as clinical notes, patient narratives, and interview transcripts are often not organized by structured variables~\cite{adnan2020role,dawson2017role,tayefi2021challenges}. Assessing AI-generated clinical outputs therefore goes beyond checking whether facts are correct; it also requires judging whether the output is complete, appropriate to the clinical context, and safe for its intended use~\cite{adnan2020role}.

Traditional evaluation pipelines face several challenges. Expert annotation remains the gold standard but is slow and expensive, which restricts both dataset size and evaluation throughput~\cite{gu2024survey, malmasi2018extracting}. Automated metrics (e.g., BLEU~\cite{papineni2002bleu}, ROUGE~\cite{lin2004rouge}, BERTscore~\cite{zhang2019bertscore}) measure lexical similarity rather than medical correctness or reasoning validity. Structured benchmarks often simplify clinical tasks into fixed-answer formats that do not fully capture ambiguity or contextual variation in real-world practice. This limitation is particularly important in healthcare, where many quality indicators are embedded in narrative clinical notes rather than structured fields~\cite{malmasi2018extracting}. These challenges have motivated growing interest in scalable evaluation methods that can approximate expert judgment.

A growing response to these gaps is ``LLM-as-a-Judge,'' which uses an LLM to score another system's output against task-specific criteria such as factual accuracy, completeness, safety, or clinical appropriateness~\cite{gu2024survey, zheng2023judging}. Prior work suggests that LLM-based judges can approximate human judgments in reasoning, summarization, and dialogue tasks~\cite{kobayashi2024large, pan2024human}. In healthcare, they have been used to evaluate clinical note generation~\cite{croxford2025evaluating}, diagnostic reasoning~\cite{sarvari2025rapidly}, medical question answering~\cite{zhao2026automating}, and radiology report summarization~\cite{vasilev2025evaluating}. Their advantages lie in the capacity to scale evaluation across large datasets while supporting multidimensional assessment of clinical quality, beyond what can be captured by simple numerical scores.

However, applying LLM-as-a-Judge to clinical tasks raises important concerns. Clinical evaluation requires domain expertise and careful consideration of patient safety. In this regard, LLM judges may hallucinate, overestimate reasoning quality, exhibit bias, or share failure modes with the models they evaluate~\cite{asgari2025framework, zack2024assessing}. Such errors may have substantial downstream consequences if they shape deployment decisions~\cite{kim2025medical}. These risks also indicate that LLM applications in healthcare require continuous monitoring rather than one-time evaluation, making scalable automated evaluation essential. However, existing reviews are largely situated in computer science and focus on general-purpose LLM evaluation~\cite{li2024llms, li2025generation, gu2024survey}. To the best of our knowledge, only one prior review has examined a closely related topic~\cite{li2026scoping}; however, it primarily focuses on risk validation and governance frameworks rather than systematically characterizing LLM-as-a-Judge applications or judge–human alignment across healthcare tasks. These gaps motivate the following research questions.

\vspace{2mm}
\begin{itemize}[topsep=0pt, itemsep=0pt, leftmargin=*]
    \item \textbf{RQ1.} How has LLM-as-a-Judge been applied across healthcare and clinical tasks?
    \item \textbf{RQ2.} What LLMs and technical approaches have been used to implement LLM-as-a-Judge?
    \item \textbf{RQ3.} What measures have been used to evaluate judge-human agreement and to what extent do LLM judges align with human experts?
    \item \textbf{RQ4.} What are the key opportunities and failure modes of LLM-as-a-Judge in clinical contexts?
\end{itemize}
\vspace{2mm}

In this review, we examine existing literature on the ``LLM-as-a-Judge'' method in healthcare and clinical applications. We first characterize how LLM judges have been applied, with attention to application areas such as clinical decision support, clinical natural language processing (NLP), medical knowledge \& question answering (QA), and medical communication. We then examine the judge models and technical approaches, including prompt engineering, ensemble, retrieval-augmented generation (RAG), fine-tuning, and multi-agent frameworks. We further summarize empirical evidence on judge-human alignment, focusing on commonly reported metrics such as agreement rate, Cohen's $\kappa$, and correlation with human annotation. Finally, we identify key opportunities and failure modes of LLM-as-a-Judge in clinical contexts, including risks related to hallucination, bias, and insufficient human validation. Together, this study provides a comprehensive review of where LLM-as-a-Judge is being used in healthcare, how it is implemented, how its reliability is evaluated, and where safeguards remain necessary.
\section{DATA \& METHODS}
\label{sec:methods}

\subsection{Data Preparation}

\begin{figure*}[ht]
    \centering
    \includegraphics[width=0.8\linewidth]{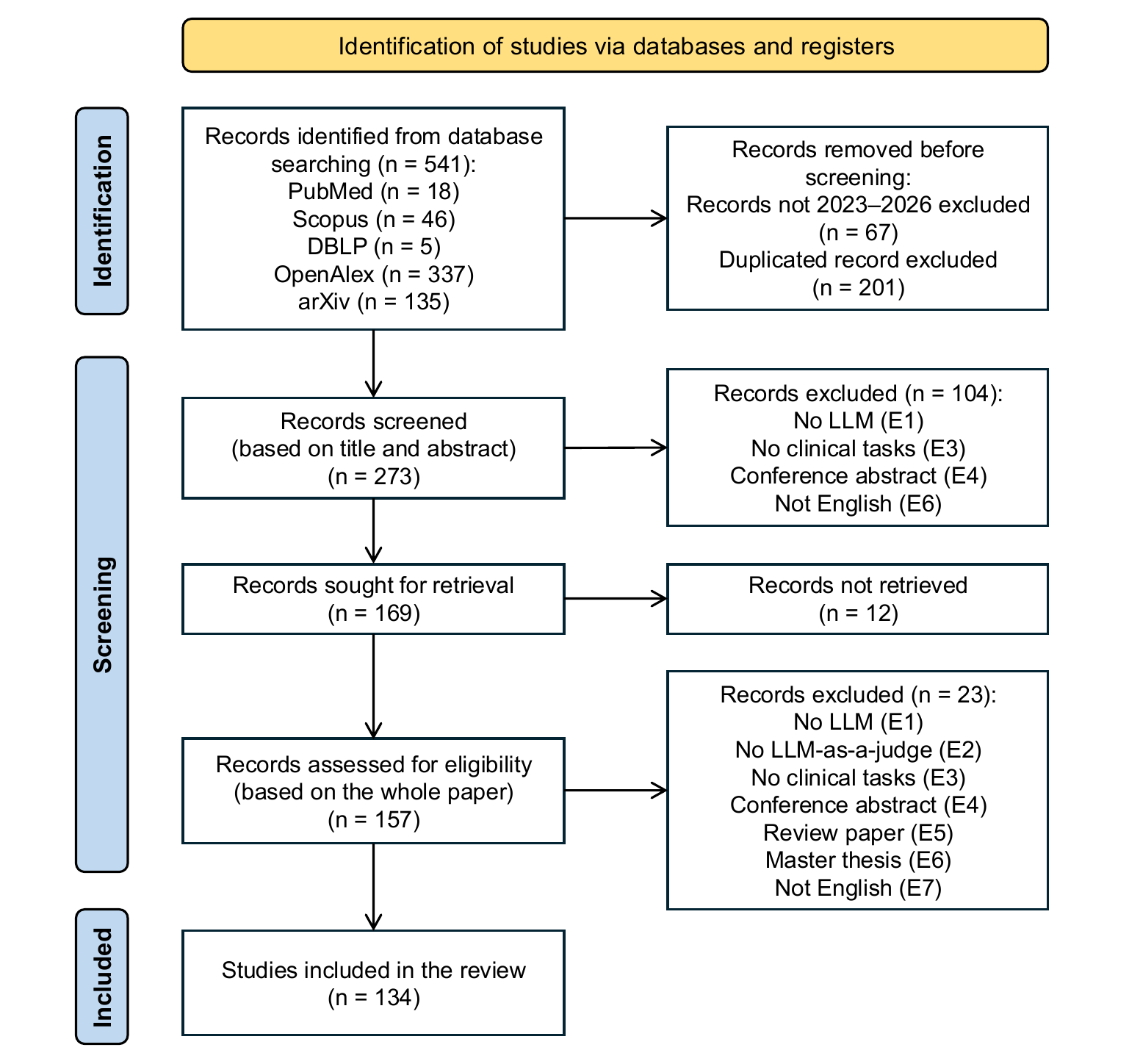}
    \caption{PRISMA flow diagram illustrating the study filtering process.}
    \label{fig:prisma}
\end{figure*}

The study begins with data preparation and study filtering. We conduct a systematic literature search in accordance with PRISMA guidelines~\cite{page2021prisma, page2021updating} across five academic databases, including PubMed\footnote{PubMed: \href{https://pubmed.ncbi.nlm.nih.gov}{https://pubmed.ncbi.nlm.nih.gov}.}, Scopus\footnote{Scopus: \href{https://www.scopus.com/pages/home}{https://www.scopus.com/pages/home}.}, DBLP\footnote{DBLP: \href{https://dblp.org}{https://dblp.org}.}, OpenAlex\footnote{OpenAlex: \href{https://openalex.org}{https://openalex.org}.}, and arXiv\footnote{ArXiv: \href{https://arxiv.org}{https://arxiv.org}.}. The search covers publications from January 2023 to February 2026 and is performed on a paper's title and abstract. The search strategy and inclusion/exclusion criteria are summarized in Table~\ref{tab:search_strategy}, and the study selection process is illustrated in Figure~\ref{fig:prisma}. 

\begin{table*}[ht]
\centering
\footnotesize
\caption{Literature search strategy and selection criteria.}
\label{tab:search_strategy}
    \begin{tabular}{p{2.8cm}p{10.2cm}}
    \toprule
    \textbf{Component} & \textbf{Description} \\
    \midrule
    Search Databases & PubMed, Scopus, DBLP, OpenAlex, arXiv \\
    \midrule
    Search Date Range & January 1, 2023 -- February 28, 2026 \\
    \midrule
    Search Terms & (``LLM-as-a-Judge'' OR ``LLM as a judge'' OR ``LLM as judge'' OR ``large language model as judge'' OR ``large language model as a judge'' OR ``agent as a judge'' OR ``agent as judge'' OR ``GPT as judge'' OR ``GPT as a judge'' OR ``Gemini as judge'' OR ``Gemini as a judge'' OR ``Claude as judge'' OR ``Claude as a judge'' OR ``Llama as judge'' OR ``Llama as a judge'' OR ``Qwen as judge'' OR ``Qwen as a judge'' OR ``DeepSeek as judge'' OR ``DeepSeek as a judge'')
    
    \textbf{AND} 
    (``health'' OR ``healthcare'' OR ``medical'' OR ``clinical'' OR ``diagnosis'' OR ``diagnostic'' OR ``patient'' OR ``EHR'' OR ``electronic health record'' OR ``biomedical'' OR ``medicine'') \\
    
    \midrule
    Screening Criteria & \textbf{Inclusion:} (I1) Uses LLM; (I2) Employs LLM as judge; (I3) Addresses clinical/healthcare tasks; (I4) Original research; (I5) Written in English. 
    
    \textbf{Exclusion:} (E1) Not using LLM; (E2) Not using LLM as a judge; (E3) No clinical/healthcare relevance; (E4) Conference abstract; (E5) Review or survey paper; (E6) Master thesis; and (E7) Non-English publications \\
    \bottomrule
\end{tabular}
\end{table*}

The screening process is conducted in multiple stages. First, filtering by publication year (2023–2026) reduces the dataset to 474 records. After exact duplicates are removed, 287 records remain. A subsequent similarity analysis based on author names and titles identifies and excludes 14 additional duplicate records, primarily cases in which official publications duplicate earlier preprint versions. This process results in 273 unique records. Title and abstract screening is then performed according to the predefined inclusion criteria, with a specific focus on studies using an LLM-as-a-Judge methodology in healthcare contexts. This step yields 169 potentially relevant papers for full-text review, of which 157 full-text articles are retrieved (12 excluded because the full-text PDFs could not be accessed). Full-text screening is conducted by two authors using the predefined eligibility criteria in Table~\ref{tab:search_strategy}, including use of an LLM as a judge or evaluator, relevance to clinical or healthcare tasks, original research papers, and English-language publication. This process results in 134 studies meeting all criteria for final inclusion.

\subsection{Data Extraction and Coding}

For each of the 134 studies retained after full-text screening, we extract structured information using a pre-specified codebook of 18 key fields, organized into four groups: (i) bibliographic metadata, (ii) study and clinical context, (iii) judge configuration, and (iv) evaluation and validation. Fields are coded with controlled vocabularies or short free-text narratives, depending on whether a closed taxonomy is practical. The complete codebook with field-level definitions and example values is provided in Appendix Table~\ref{tab:codebook}.

The first group records bibliographic metadata: paper title, author list, author affiliations, publication year-month, and DOI. The second group captures study and clinical context, including the study dataset, data modality (e.g., text, images), and clinical context. Clinical context is coded as one of five top-level scenarios, including \textit{Clinical Decision Support}, \textit{Clinical NLP}, \textit{Medical Knowledge \& QA}, \textit{Medical Education}, and \textit{Other}, paired with a free-text task description. The taxonomy is iteratively refined to be mutually exclusive at the scenario level while preserving task-level granularity for downstream analysis.

The third group describes judge configuration. We record the judge content (what type of output is evaluated) together with the specific evaluation dimensions, specific LLM used as a judge with version and provider, and the techniques employed. The technique vocabulary uses non-mutually-exclusive labels, including \textit{Prompt Engineering}, \textit{Ensemble}, \textit{RAG}, \textit{Calibration}, \textit{Fine-tuning}, \textit{Multi-agent}, \textit{Distillation}, as many combine two or more. The fourth group records how each study assesses its judge. We capture the list of reported evaluation metrics (e.g., percent agreement, Cohen's $\kappa$, Pearson's $r$, F1, ROC-AUC), corresponding numerical performance, and validation sample size. A binary \textbf{judge against human validation} field flags whether the judge has been compared against expert human ratings.

\subsection{LLM-as-a-Judge Framework in Healthcare}
\label{sec:judge_framework}

\begin{figure}[htbp]
    \centering
    \includegraphics[width=\linewidth]{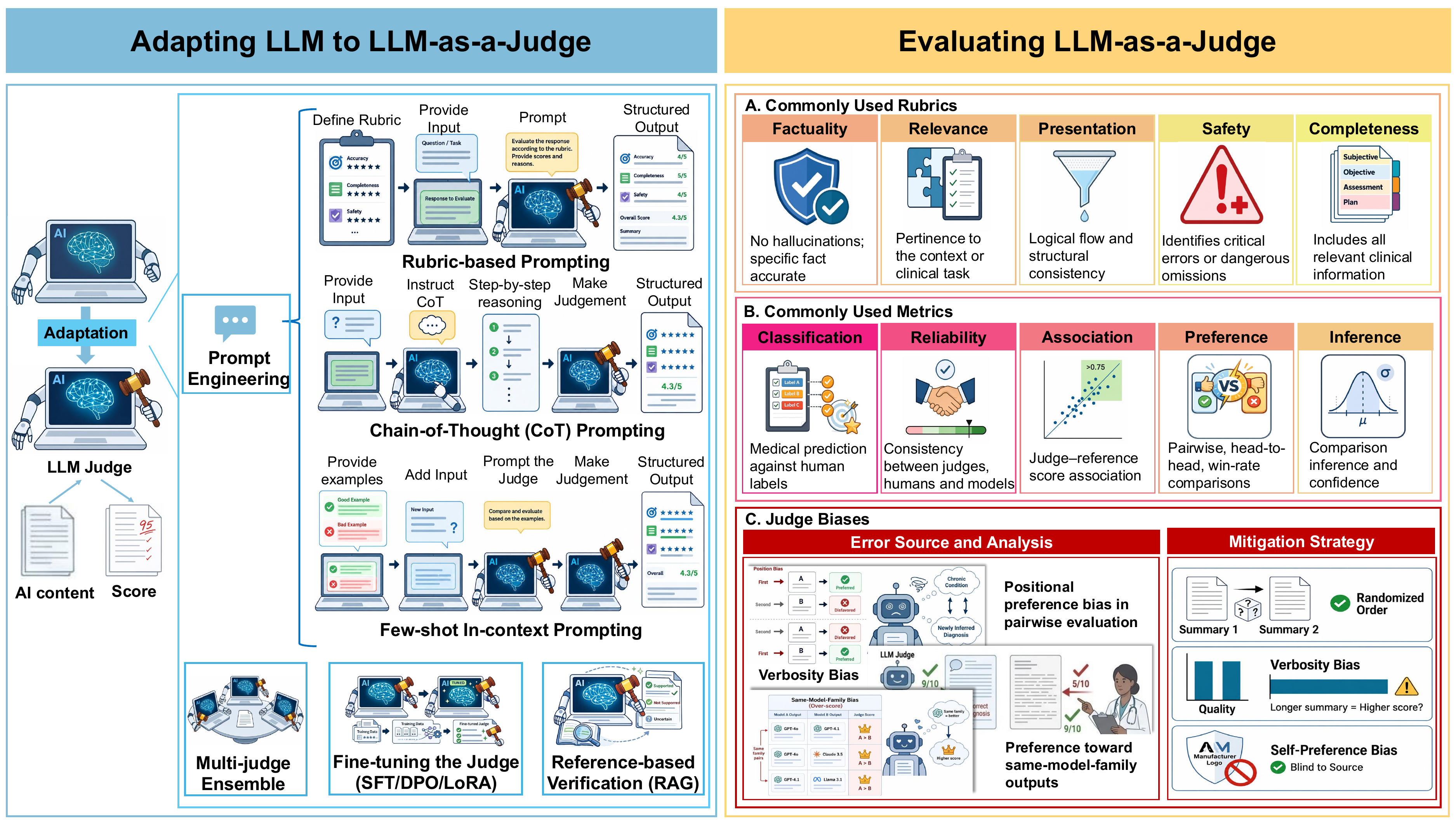}
    \caption{\textbf{Overview of LLM-as-a-Judge frameworks in healthcare.} The framework typically includes adaptation strategies, evaluation rubrics and metrics, and common judge biases with corresponding mitigation approaches.}
    \label{fig:framework}
\end{figure}

LLM-as-a-Judge refers to the use of an LLM to evaluate outputs generated by another model or system~\cite{gu2024survey,zheng2023judging}. In healthcare, this approach often cannot be treated as a simple automated scoring step. Clinical outputs often require interpretation in context, comparison with evidence, and attention to patient safety. We therefore organize its use in healthcare into three parts: (i) defining the healthcare task, (ii) configuring the judge, and (iii) validating the resulting judgments. After the healthcare task is defined, the judging process typically centers on the interaction between judge configuration and evaluation design, as illustrated in Figure~\ref{fig:framework}. Specific technical implementations of LLM-as-a-Judge methods are discussed in Section~\ref{sec:judge_techniques}.

\textbf{Step 1: Task and rubric specification.} The first step defines the healthcare task under evaluation and the judge's task with respect to it. We identify and group current studies into five categories: \textit{Clinical Decision Support} (outputs related to diagnostic, prognostic, or therapeutic decision-making), \textit{Clinical NLP} (note/discharge summary generation for clinical documentation and clinical entity extraction for information structuring), \textit{Medical Knowledge \& QA} (factual and reasoning-based responses to biomedical or clinical questions), \textit{Medical Communication} (applications involving communication between healthcare systems, patients, and healthcare professionals), and \textit{other biomedical topics}. Then the judge's task may take several forms: scoring a single output against a rubric~\cite{Cheng2026scaling, vasilev2025evaluating, croxford2025evaluating}, comparing two candidates head-to-head~\cite{hosseini2024benchmark, morse2025custom}, checking factual consistency against a reference~\cite{chung2025verifact, steinigen2026fact}, or assigning a categorical safety label~\cite{aali2025medval}. The chosen task in turn determines the rating scale and the judgment format, which may be a structured rubric-anchored judgment, a pairwise preference, or an agreement-based label aligned with an expert reference.

\textbf{Step 2: Judge configuration and enhancement.} A general-purpose LLM is configured for the task defined in Step~1 and, where needed, augmented with additional information or structure. The base configuration encodes the rubric directly into the judge through prompt-level mechanisms such as rubric-based prompting~\cite{Cheng2026scaling, vasilev2025evaluating, croxford2025evaluating}, chain-of-thought (CoT) instructions~\cite{cai2025exploring, wu2025chain}, or few-shot exemplars \cite{wu2025automated}. On top of this, several extension strategies are available. The judge model can be fine tuned via supervised fine-tuning~\cite{zheng2025llm, laskar2025improving}, parameter-efficient methods~\cite{hu2022lora}, preference optimization~\cite{rafailov2023direct}, or distillation from a stronger teacher~\cite{aali2025medval, yao2026medqa}. The judge can also be provided with retrieved external material (e.g., guidelines, drug references, or EHR snippets) so that factuality is checked against a ground source~\cite{sarvari2025rapidly, yan2026livemedbench}. Multiple judges can also be deployed to vote in an ensemble or interact in a multi-agent configuration~\cite{chen2025multiagent, chen2025gaps}. These extensions are not mutually exclusive, and the choice should follow the clinical risk profile and rubric structure.

\textbf{Step 3: Validation and bias mitigation.} The final step is to validate the judge's performance, often through the comparison with human or expert assessment. The validation metric depends on the form of the judgment. Categorical and ordinal ratings are commonly assessed with Cohen's $\kappa$ or ICC, continuous scores with Pearson's $r$ or Spearman's $\rho$, classification labels with standard classification metrics, and pairwise judgments with win rates. Additional mitigation strategy is also needed because LLM judges can introduce their own biases or errors~\cite{dai2025model, laskar2025improving, vasilev2025evaluating, williams2025human, wu2025automated}. A judge may prefer the first answer in a pair, favor longer responses, overvalue fluent but weak reasoning, or show preference for outputs from related models. These biases can be reduced by randomizing response order, blinding the source model, controlling response length, repeating the judgment, or aggregating results across runs or judges.

\subsection{LLM-as-a-Judge Techniques}
\label{sec:judge_techniques}

To describe how LLM-as-a-Judge systems are implemented in healthcare studies, we identify the following major technical strategies used to shape the judge's behavior, as illustrated in Figure~\ref{fig:framework} (blue box). These strategies are not mutually exclusive. A single study may, for example, use a rubric-based prompt and aggregate judgments from several models.

\textbf{Prompt engineering} is the most typical mechanism for instructing the judge. It uses task instructions, scoring criteria, in-context learning examples, and response-format constraints (e.g., JSON)~\cite{marvin2023prompt}. In healthcare studies, prompt engineering often appears as rubric-based evaluation, where the prompt can define dimensions such as factual accuracy, completeness, safety, empathy, or clinical usefulness~\cite{he2024quality, khatwani2025brittleness, sayeedi2026route, jarchow2025benchmarking}. Some prompts use Likert scales~\cite{cardenal2025hivmedqa}, while others ask for binary or categorical judgments, such as correct versus incorrect~\cite{sesen2025development}. Few-shot examples can be added to illustrate the expected judgment standard~\cite{sangwon2025evaluating}. CoT instructions may also be used when the study asks the judge to explain its reasoning before assigning a final score~\cite{cai2025exploring}.

\textbf{Ensemble} methods combine more than one judgment into a final score or label. The component judgments may come from different LLMs, repeated runs of the same LLM, or alternative prompts. Majority voting is commonly used for categorical labels, whereas numerical scores are usually averaged or combined through weighted aggregation. Ensemble designs are intended to reduce dependence on a single model and to mitigate model-specific biases, such as verbosity preference or self-preference~\cite{chen2025multiagent, chen2025gaps}.

\textbf{Retrieval-augmented generation (RAG)} grounds models in external evidence before the evaluation is made~\cite{lewis2020retrieval}. In the study context, retrieval supports the judging process rather than the generation of the candidate answer. The retrieved material can include clinical guidelines, institutional protocols, drug references, electronic health record (EHR) snippets, biomedical knowledge bases, or web-search results. The judge then uses this evidence to assess factuality and evaluate consistency with other models' outputs~\cite{sarvari2025rapidly, yan2026livemedbench}. We treat RAG as an LLM-as-a-Judge technique only when the retrieved evidence is supplied to the judge itself, rather than only to the model whose output is being evaluated.

\textbf{Calibration} refers to procedures that reduce predictable bias or improve score stability~\cite{wang2024large}. For pairwise comparison, a common approach is to swap the order of candidate responses so that each answer appears in both positions before the final preference is determined. Other procedures include randomizing answer order, normalizing scores across repeated runs, and testing whether the judge favors longer, more confident, or self-generated responses~\cite{shi2025judging, li2025calibraeval}. Unlike general prompt design, calibration focuses on making the judge's scores more comparable and less sensitive to irrelevant factors such as answer order, length, or model identity.

\textbf{Fine-tuning} adapts the judge model using evaluation-specific data~\cite{chiang2025tract}. This may involve supervised fine-tuning on expert-labeled rubric scores~\cite{zheng2025llm, laskar2025improving}, parameter-efficient adaptation such as LoRA~\cite{hu2022lora}, or preference optimization~\cite{rafailov2023direct} using human or model-generated judgments. \textbf{Distillation} is a related but narrower strategy in which a smaller judge is trained to approximate the outputs of a stronger teacher model~\cite{aali2025medval,yao2026medqa}. Fine-tuning mainly adapts the judge to a domain or task, whereas distillation transfers judgment behavior to a cheaper or more deployable evaluator.

\textbf{Multi-agent} judging uses several LLM agents that interact before a final evaluation is produced~\cite{chen2025multi}. Unlike a standard ensemble, these agents are not simply independent voters. They may assume different roles, critique one another, debate alternative interpretations, or pass intermediate assessments to a reviewer or supervisor~\cite{luo2025dialogguard, wu2025towards, chen2025multi}. This design is often used for multidimensional or safety-sensitive evaluations, where different agents can focus on factual consistency, patient profile analysis, or clinical reasoning.

Together, these techniques define the main ways in which LLM judges are configured for evaluation across healthcare tasks. For each study, we record the technique category and a brief implementation description. Table~\ref{tab:judge_techniques} summarizes how each technique is implemented and provides representative examples from the reviewed studies.

\begin{table*}[ht]
\centering
\footnotesize
\caption{Major LLM-as-a-Judge techniques and their implementation in healthcare studies.}
\label{tab:judge_techniques}
    \begin{tabular}{p{0.12\linewidth} p{0.23\linewidth} p{0.35\linewidth} p{0.20\linewidth}}
    \hline
    \textbf{Technique} & \textbf{Purpose} & \textbf{Typical implementation} & \textbf{Example studies} \\
    \hline
    Prompt engineering
    & Guides the judge model through explicit instructions and evaluation criteria. 
    & Rubric-based scoring, anchored Likert scales, binary or categorical labels, few-shot examples, CoT reasoning, prompt optimization, and structured outputs such as JSON or key-value schemas. 
    & \citet{sangwon2025evaluating, khatwani2025brittleness, sayeedi2026route, jarchow2025benchmarking, cai2025exploring} \\
    \hline
    Ensemble 
    & Combines multiple judgments to reduce instability and model-specific bias. 
    & Majority voting for categorical labels; averaging or weighted aggregation for numerical scores; aggregation across different models, repeated runs, or prompt variants.
    & \citet{williams2025human, li2025medguide, chen2025gaps, abdullahi2026persona} \\
    \hline
    RAG 
    & Grounds the judge's evaluation in external evidence. 
    & Retrieval from clinical guidelines, institutional protocols, references, EHR documentation, biomedical knowledge bases, or web-search results before judgment. 
    & \citet{sarvari2025rapidly, yan2026livemedbench} \\
    \hline
    Calibration 
    & Reduces predictable judge biases and improves score stability. 
    & Position swapping in pairwise comparison, randomized answer order, score normalization, verbosity-bias checks, and self-preference tests. 
    & \citet{hosseini2024benchmark, chang2025multi} \\
    \hline
    Fine-tuning 
    & Adapts the judge model to a specific evaluation task or clinical domain. 
    & Supervised fine-tuning on expert labels, LoRA/QLoRA adaptation, and preference optimization using evaluation-specific data. 
    & \citet{he2026mlb, croxford2025evaluating, nghiem2025balancing} \\
    \hline
    Distillation 
    & Transfers judgment behavior from a stronger teacher model to a smaller evaluator. 
    & Teacher-generated labels, scores, or preference judgments are used to train a smaller model for efficient or local deployment. 
    & \citet{aali2025medval,yao2026medqa}  \\
    \hline
    Multi-agent
    & Uses interaction among multiple agents to evaluate complex scenarior or judge reasoning outputs.
    & Role-specialized debate, critic, such as reviewer pipelines, persona-based juries, and cross-informed voting.
    & \citet{luo2025dialogguard, wu2025towards, chen2025multi} \\
    \hline
\end{tabular}
\end{table*}

\section{RESULTS}
\label{sec:results}

\begin{figure*}[ht]
    \centering
    \includegraphics[width=0.8\linewidth]{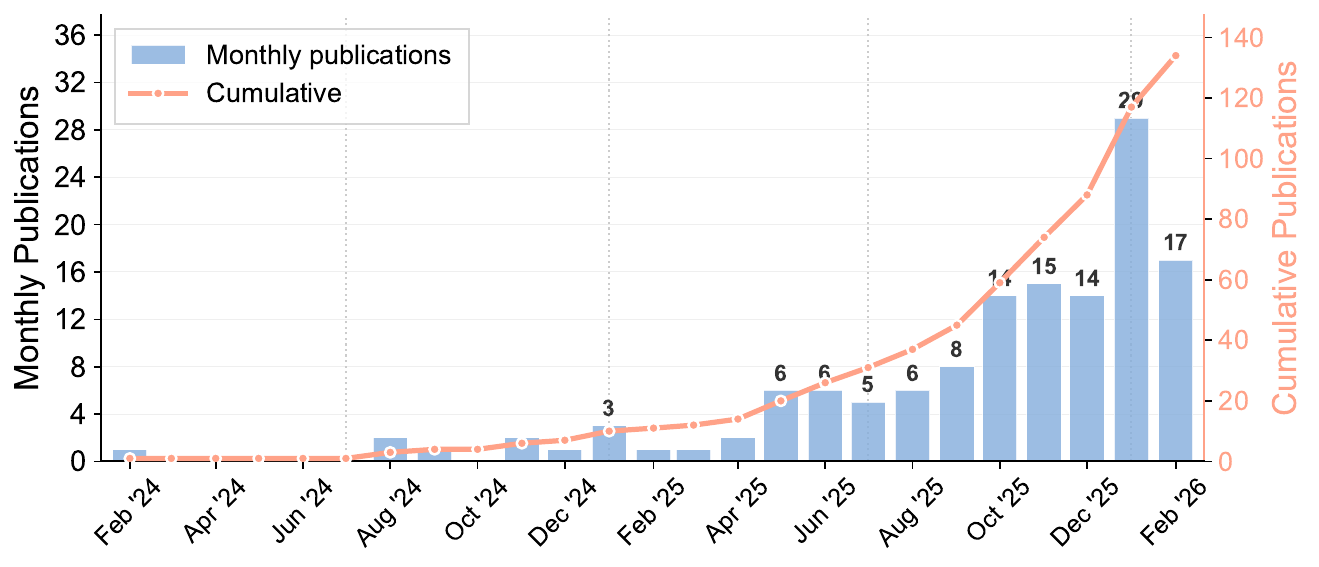}
    \caption{Publication trend of LLM-as-a-Judge studies in healthcare during the study period. Bars show the number of monthly publications, and the line shows the cumulative number of eligible publications over time. The sharp increase in 2025 and early 2026 indicates the rapid emergence of LLM-as-a-Judge in the studies.}
    \label{fig:trend}
\end{figure*}

Research on LLM-as-a-Judge in healthcare expanded rapidly over the two years studied (Figure~\ref{fig:trend}). Only 7 studies were published in 2024, but the literature grew sharply to 81 studies in 2025, with an additional 46 studies published in the first two months of 2026 alone. This trend indicates that automated evaluation has quickly become an important methodological direction for assessing AI-generated clinical content. Additional analyses of the geographic distribution of publications and author collaborations based on institutional affiliations are presented in Appendix~\ref{app:geo_collaboration}.

The following results are organized around the study research questions. Section~\ref{sec:health_applications} addresses \textbf{RQ1} by characterizing where LLM-as-a-Judge has been applied across healthcare domains, including \textit{Clinical Decision Support}, \textit{Clinical NLP}, \textit{Medical Knowledge \& QA}, and \textit{Medical Education} (corresponding to Step~1 in Section~\ref{sec:judge_framework}). Section~\ref{sec:judge_models} addresses \textbf{RQ2} by summarizing the judge models and technical approaches used to implement LLM-based judges (corresponding to Step~2 in Section~\ref{sec:judge_framework}). Sections~\ref{sec:judge_evaluation} and ~\ref{sec:judge_alignment} address \textbf{RQ3} by mapping the evaluation metrics used across application domains and then synthesizing empirical evidence on judge-human alignment, with an illustration on agreement rate, Cohen's $\kappa$, and correlation (corresponding to Step~3 in Section~\ref{sec:judge_framework}). These results together describe the evolving applications, methodological design, and reliability evidence supporting LLM-as-a-Judge for healthcare applications.

\subsection{Healthcare Applications of LLM-as-a-Judge}
\label{sec:health_applications}

\begin{figure}[htbp]
    \centering
    \includegraphics[width=\linewidth]{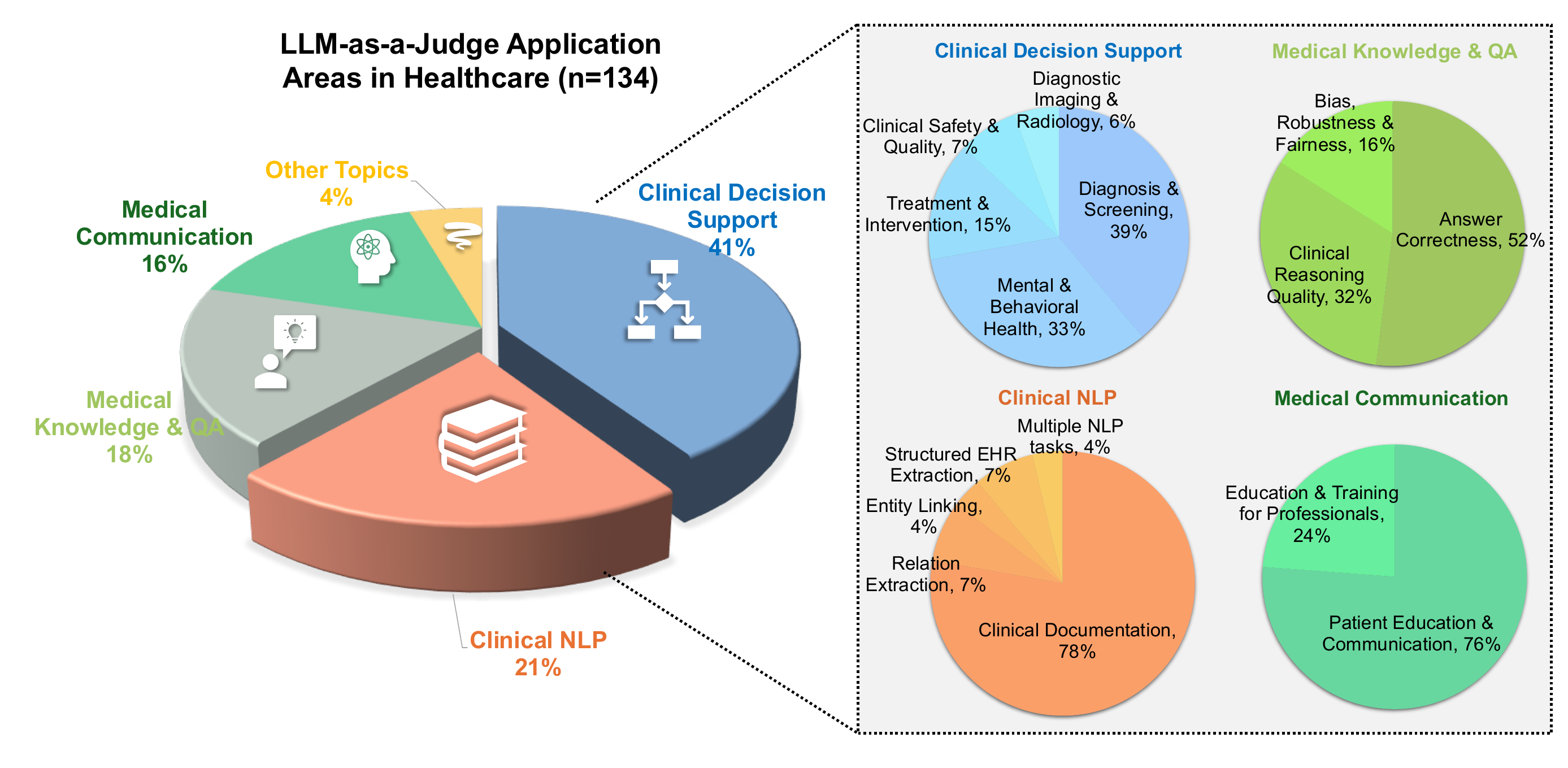}
    \caption{\textbf{LLM-as-a-Judge research areas in healthcare.} Major research domains include \textit{Clinical Decision Support}, \textit{Clinical NLP}, \textit{Medical Knowledge \& QA}, and \textit{Medical Communication}.}
    \label{fig:application_piechart}
\end{figure}

\begin{figure}[ht]
  \centering
  \resizebox{\textwidth}{!}{\input{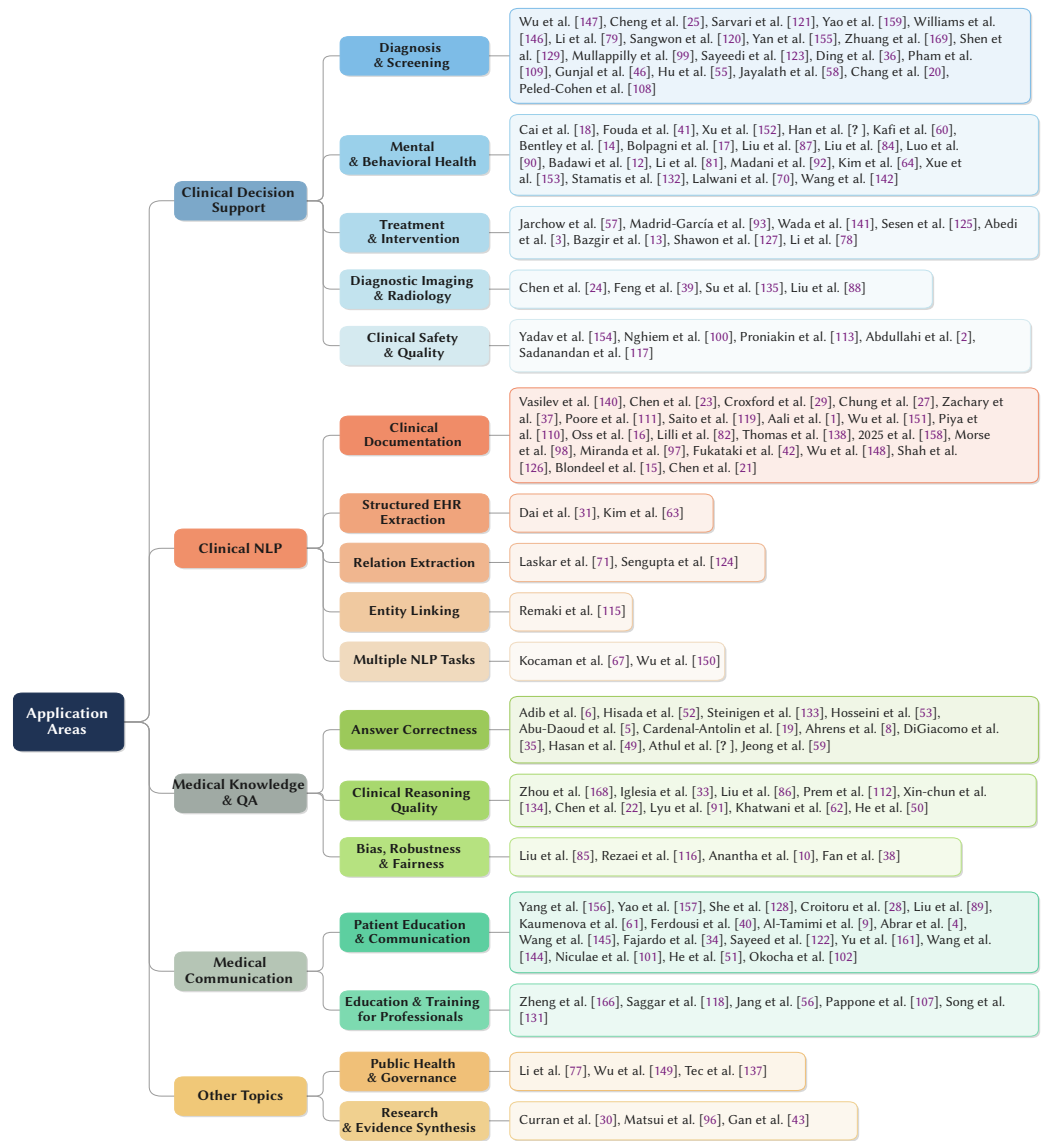}}
  \caption{\textbf{Detailed categorization of LLM-as-a-Judge research in healthcare.} Included studies are grouped according to major healthcare research domains and their corresponding application tasks.}
  \label{fig:application_tree}
\end{figure}

\textbf{Overall distribution.}
Across the 134 included studies, applications of LLM-as-a-Judge in healthcare are concentrated in \textit{Clinical Decision Support} (n=54, 40.3\%), followed by \textit{Clinical NLP} (n=28, 20.9\%), \textit{Medical Knowledge \& QA} (n=24, 17.9\%), and \textit{Medical Communication} (n=22, 16.4\%) (Figure~\ref{fig:application_piechart}). This distribution shows that LLM-as-a-Judge is primarily used in decision-critical and text-intensive settings, where scalable and consistent evaluation is needed. A more detailed task-level breakdown is shown in Figure~\ref{fig:application_tree}, which groups each study into specific healthcare applications.

\textbf{Clinical Decision Support.}
Within \textit{Clinical Decision Support}, most studies focus on diagnosis and screening tasks (n=19, 35.2\%), reflecting the use of LLM-as-a-Judge to assess diagnostic accuracy and triage decisions. For example, \citet{sarvari2025rapidly} use an LLM-as-a-Judge framework to benchmark diagnostic performance across 18 LLMs using MIMIC-IV admissions. \citet{wu2025automated} evaluate concordance between AI-generated and specialist responses in physician-to-physician eConsult scenarios. \citet{sangwon2025evaluating} use \textit{GPT-4o} as an LLM judge to assess diagnostic equivalence in free-response clinical reasoning tasks within a multi-agent, multi-turn conversational benchmarking framework.

Mental and behavioral health represents another major area within \textit{Clinical Decision Support} (n=18, 33.3\%). These studies extend evaluation beyond correctness to appropriateness, cultural sensitivity, empathy, and safety in emotionally sensitive settings. For example, \citet{liu2026tailored} use LLM judges alongside human annotators and clinical psychologists to evaluate cultural sensitivity and emotional appropriateness across diverse populations. \citet{han2026optimizing} leverage LLM judges within a two-stage evaluation framework for culturally adapted mental health counseling models with psychiatrist-based expert validation. \citet{wang2025chatthero} apply LLM-as-a-Judge alongside blinded clinicians to assess empathy, clinical relevance, and behavioral change effectiveness in a multi-session language agent for substance use disorder recovery.

By contrast, treatment-related evaluations (n=8, 14.8\%) and clinical safety and quality assessment (n=5, 9.3\%) remain underrepresented. This imbalance suggests that current research gives greater attention to front-end decision-making, such as diagnosis and triage, than to downstream intervention validation and safety monitoring, despite the importance of these areas for real-world clinical deployment.

\textbf{Clinical NLP}.
In \textit{Clinical NLP} tasks (n=28), LLM-as-a-Judge is mainly used for clinical documentation tasks (n=21, 75.0\%), reflecting a strong focus on evaluating generated and summarized clinical notes. \citet{vasilev2025evaluating} assess EHR-based summaries across multiple quality dimensions. \citet{croxford2025evaluating} report strong agreement between LLM judges and human evaluators in multi-document summarization tasks. \citet{saito2025generation} employ an LLM-as-a-Judge framework to evaluate automatically generated SOAP-format nursing records, with emphasis on hallucination and information omission.

Other \textit{Clinical NLP} applications, including relation extraction (n=2, 7.1\%), entity linking (n=1, 3.6\%), and structured data extraction (n=2, 7.1\%), are less common. This pattern suggests that LLM-as-a-Judge is most often applied to holistic, language-intensive evaluation tasks, where traditional rule-based or metric-based methods may be insufficient. Fine-grained structured prediction tasks remain less explored, potentially because standardized rubrics for LLM-based judging are less developed in these settings.

\textbf{Medical Knowledge \& QA}.
In \textit{Medical Knowledge \& QA} tasks (n=24), evaluation primarily focuses on answer correctness (n=11, 45.8\%), followed by clinical reasoning quality (n=9, 37.5\%) and bias, robustness, and fairness-related dimensions (n=4, 16.7\%). This pattern suggests a shift from outcome-oriented evaluation of factual correctness toward process-aware assessment of reasoning quality, although fairness and robustness remain relatively underexplored. For example, \citet{hosseini2024benchmark} use LLM judges to evaluate long-form medical QA responses for correctness, helpfulness, harmfulness, efficiency, and bias alignment with physician annotations. \citet{zhou2025automating} and \citet{liu2026closing} apply LLM-as-a-Judge frameworks to assess the quality and fidelity of step-by-step clinical reasoning. \citet{liu2024decoy} employ LLM judges to examine cognitive bias and robustness in COVID-19 misinformation assessment under decoy-effect settings.

\textbf{Medical Communication.}
In \textit{Medical Communication} applications (n=22), most studies focus on patient education and communication (n=17, 77.3\%), while fewer studies address education and training for healthcare professionals (n=5, 22.7\%). This distribution highlights the predominant use of LLM-as-a-Judge frameworks in evaluating patient-facing interactions. For example, \citet{abrar2025empirical} use cross-model LLM judges to evaluate the quality and expert alignment of consumer-facing medical question-answering responses using real-world Reddit-based health queries. \citet{saggar2026ai} use an LLM-as-a-Judge framework to assess the educational utility of simulated clinician--patient interactions in pediatric OSCE-style training scenarios.

\textbf{Other areas.}
A small subset of studies (n=6) applies LLM-as-a-Judge to public health and governance \cite{li2026scaling, wu2025beyond, tec2025rule}, as well as research and evidence synthesis tasks \cite{curran2024examining, matsui2024human, gan2025polyrag}. Although limited in number, these studies indicate the potential value of LLM-based evaluation for system-level decision-making and policy evaluation. Their limited representation also suggests that the field has not yet fully addressed macro-level healthcare challenges, including regulatory evaluation and population health assessment.

\subsection{LLM-as-a-Judge Models and Techniques}
\label{sec:judge_models}

\begin{figure*}[ht]
    \centering
    \includegraphics[width=\linewidth]{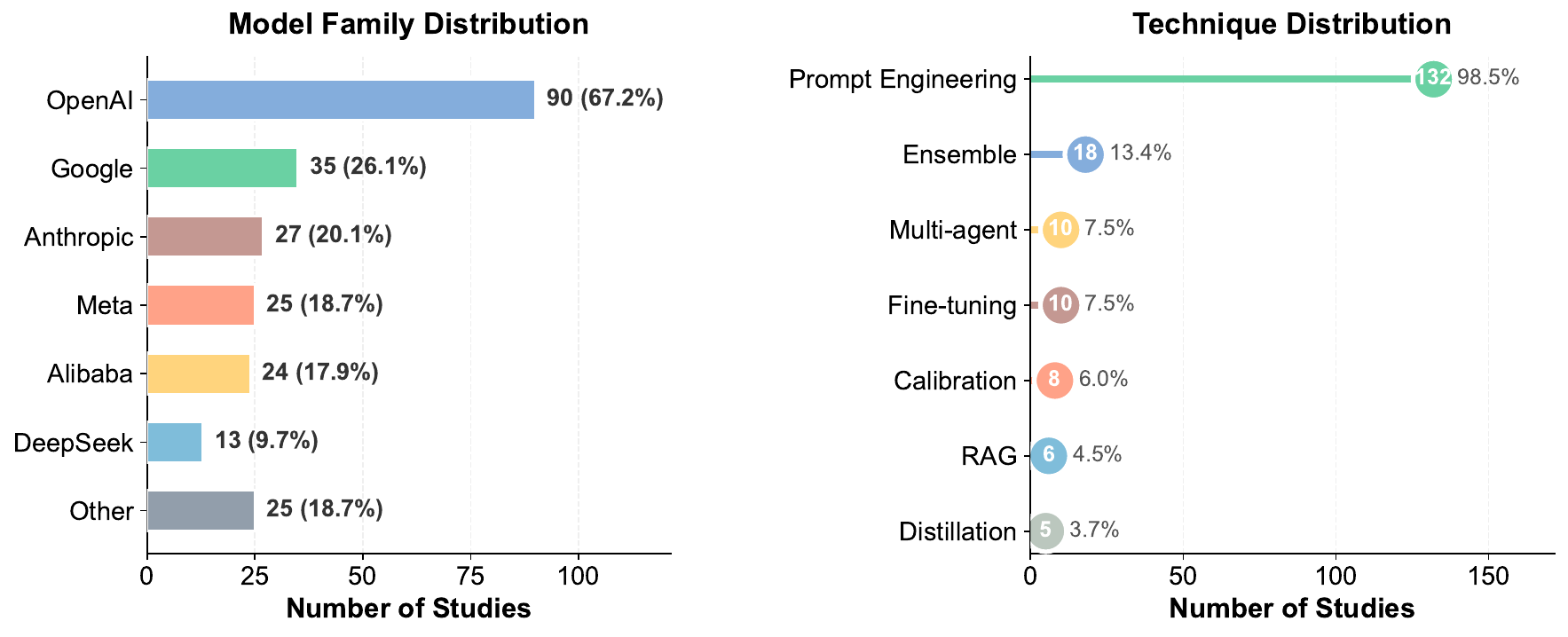}
    \caption{\textbf{Model family and technique distributions for LLM-as-a-Judge in healthcare based on N=134 studies.} (a) Number of studies employing each model family, counted at the study level. (b) Distribution of technical approaches for configuring LLM judges. Techniques are not mutually exclusive.}
    \label{fig:techniques}
\end{figure*}

\textbf{Overall distribution.}
Figure~\ref{fig:techniques} presents the overall distribution of judge model families and technical approaches. Regarding the judge models, OpenAI models are the most frequently used judges, appearing in 90 of 134 studies (67.2\%), followed by Google (\textit{Gemini}, \textit{Gemma}; n=35, 26.1\%), Anthropic (\textit{Claude}; n=27, 20.1\%), Meta (\textit{LLaMA}; n=25, 18.7\%), Alibaba (\textit{Qwen}; n=24, 17.9\%), and DeepSeek (n=13, 9.7\%) (Figure~\ref{fig:techniques}a). Most studies employ models from multiple families, consistent with multi-model jury configurations designed to reduce provider-specific biases. 

Regarding the judge techniques, prompt engineering is used in 132 of 134 studies (98.5\%) (Figure~\ref{fig:techniques}b). Common strategies include rubric-based prompting, few-shot prompting, and CoT prompting. Beyond prompting, ensemble methods combining multiple judges are used in 18 studies (13.4\%), multi-agent configurations in 10 studies (7.5\%), and calibration in 8 studies (6.0\%). Fine-tuning via supervised learning or parameter-efficient adaptation appears in 10 studies (7.5\%), RAG in 6 studies (4.5\%), and knowledge distillation in 5 studies (3.7\%). 

\begin{figure*}[ht]
    \centering
    \includegraphics[width=\linewidth]{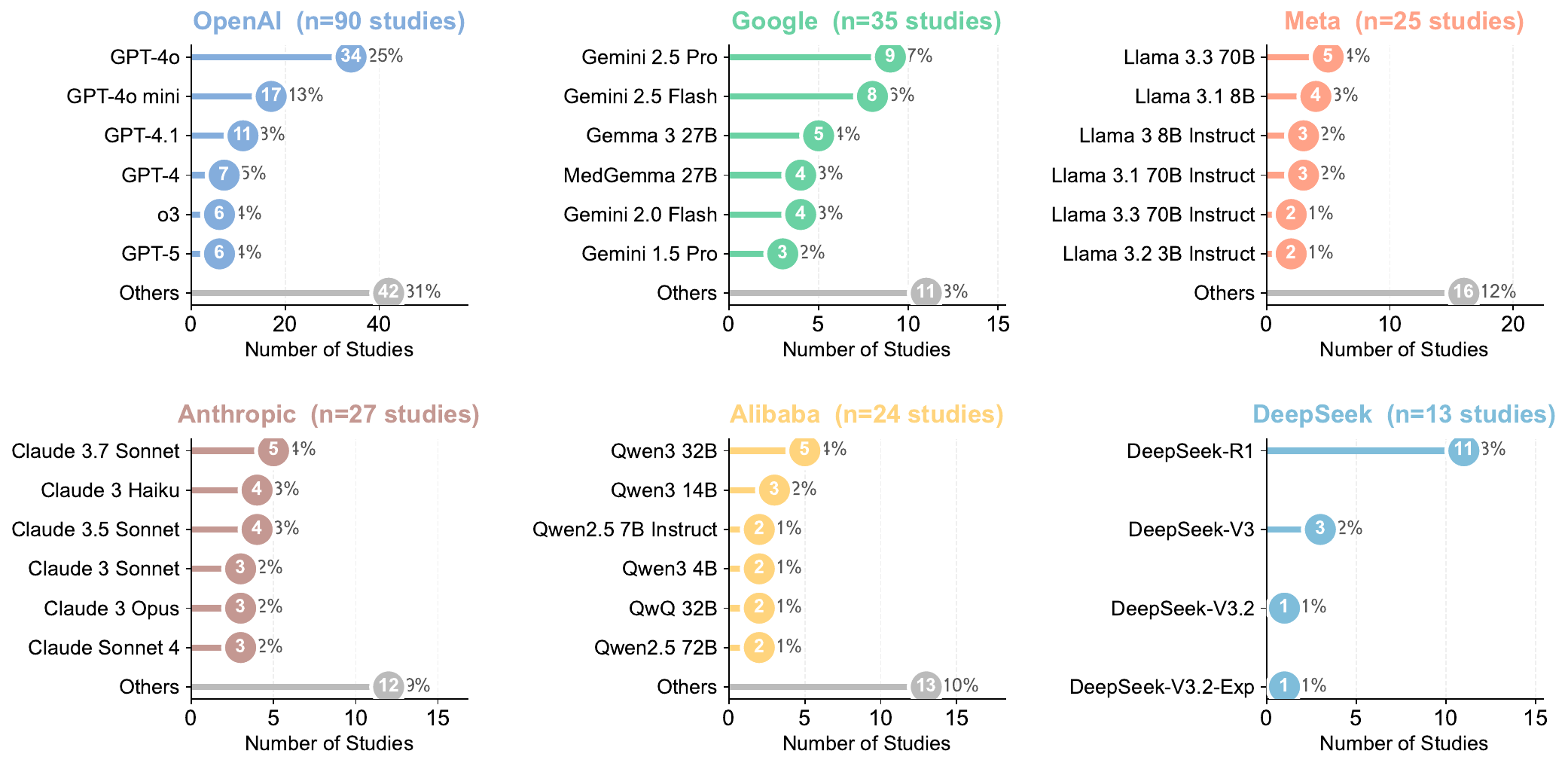}
    \caption{\textbf{Specific model versions used as LLM judges.} Each subplot shows the top most frequently used models within each major model family: OpenAI (n=90 studies), Google (n=35), Meta (n=25), Anthropic (n=27), Alibaba (n=24), and DeepSeek (n=13). Numbers on each node indicate the count of studies employing that model version; percentages reflect the proportion of total included studies (N=134). ``Others'' aggregates remaining models within each family.}
    \label{fig:models}
\end{figure*}

\textbf{Judge model selection.} 
Figure~\ref{fig:models} provides a more detailed breakdown of specific model versions within each major family. Overall, closed-source models appear to be selected as judges more frequently than open-source models. Within the OpenAI family, \textit{GPT-4o} is the most common judge (n=34, 25\%), followed by \textit{GPT-4o-mini} (n=17, 13\%) and \textit{GPT-4.1} (n=11, 8\%); other OpenAI variants, including \textit{GPT-4}, \textit{o3}, and \textit{GPT-5}, are used in 31\% of studies that employ OpenAI models (Figure~\ref{fig:models}). Among Google models, \textit{Gemini-2.5-Pro} (n=9) and \textit{Gemini-2.5-Flash} (n=8) are the most prevalent, with \textit{Gemma-3-27B} (n=5) and \textit{MedGemma-27B} (n=4) also represented. For Anthropic, \textit{Claude-3.7-Sonnet} (n=5), \textit{Claude-3-Haiku} (n=4), and \textit{Claude-3.5-Sonnet} (n=4) are the most frequently used models. Among open-source model families, \textit{LLaMA-3.3-70B} (n=5) and \textit{LLaMA-3.1-8B} (n=4) are the leading Meta models, while \textit{Qwen3-32B} (n=5) is the most common Alibaba model. \textit{DeepSeek-R1} appears in 11 studies (8\%), making it the third most frequently used individual judge model overall. The temporal trend of open-source and closed-source models is presented in Appendix~\ref{app:model_temporal}.

\textbf{Technical approaches.}
For prompting strategies, rubric-based prompting is predominant, which encodes structured evaluation criteria with explicit score-level descriptors. Rubrics can range from general clinical-quality dimensions~\cite{li2025medguide} to instrument-aligned frameworks, such as PDSQI-9 for documentation quality~\cite{li2026scaling} and OSCE-style scoring for clinical examination skills~\cite{yao2026medqa}. Few-shot prompting is often used to guide model outputs; for example, eConsult concordance evaluation embeds one concordant and one discordant case within the prompt~\cite{wu2025automated}. CoT prompting is associated with improved transparency and reduced variability across evaluations~\cite{cai2025exploring, shen2025towards}. Programmatic prompt optimization also appears; GEPA-style evolutionary search is applied to optimize prompts for assessing the clinical impact of ASR errors~\cite{ellis2026unaware}.

Ensemble approaches combine outputs from multiple judges to mitigate single-model biases. Several ensemble strategies have been applied. Majority voting is used for categorical outputs. For example, three-judge voting on element-wise rubric items~\cite{chen2025gaps}. Score averaging is used for continuous outputs; for instance, MedGUIDE averages ratings from four heterogeneous judges per criterion~\cite{li2025medguide}. Weighted aggregation based on agreement with human annotations is also applied~\cite{williams2025human}. Several studies address positional bias by evaluating pairwise comparisons (in both AB and BA order) prior to voting~\cite{hosseini2024benchmark}.

Multi-agent configurations deploy multiple collaborating or adversarial agents. For example, role-specialized debate assigns agents to opposing positions; DialogGuard implements this for psychosocial safety with pro-risk and pro-safe agents and an impartial aggregator~\cite{luo2025dialogguard}. Critic-reviewer pipelines layer specialized agents (detector, critic, reviewer), each with responsibilities such as hallucination flagging or omission detection~\cite{liu2025statistically}. Persona-based juries assign each agent a stakeholder persona and conduct free debate before aggregation~\cite{chen2025multi}. Cross-informed voting, where each agent observes others' outputs before voting, achieves stronger agreement with gold-standard rankings than independent voting in de-identification model selection~\cite{wu2025towards}.

RAG provides judges with access to external knowledge sources. Among the examined studies, applications include grounding diagnoses in ABIM lab reference ranges~\cite{sarvari2025rapidly}, and triggering re-evaluation with a modified prompt when judge confidence falls below a threshold~\cite{anantha2025nanoflux}. Dual-stage factuality verification classifies atomic claims against FAISS-retrieved EHR snippets as supported, contradicted, or not found~\cite{wu2025dual}. 

For fine tuning, SFT on human-annotated rubric labels is the most common approach; for instance, a curated SFT corpus is used to fine-tune \textit{Qwen3-14B} for disputed clinical cases~\cite{he2026mlb}. QLoRA is applied in clinical summary evaluation with both SFT and direct preference optimization~\cite{croxford2025evaluating}. Knowledge distillation trains small models from more advanced teacher judges. For example, a \textit{LLaMA-3B} judge is distilled from \textit{GPT-4o-mini} safety labels~\cite{nghiem2025balancing}. MedVAL combines self-supervised distillation, consistency-based data filtering, and QLoRA fine-tuning ~\cite{aali2025medval}. Fine-tuned judges trade some alignment with frontier models for reproducibility, lower inference cost, and on-premises deployment where transmitting protected health information to external APIs is not permissible.

\subsection{LLM-as-a-Judge Performance Measures}
\label{sec:judge_evaluation}

We summarize evaluation frameworks for LLM-as-a-Judge in healthcare research along two dimensions: \emph{Rubric} and \emph{Metric} (Table~\ref{tab:evaluation}). We use \emph{Rubric} to refer to the subjective evaluation criteria adopted in prior studies when using LLM-as-a-Judge to assess newly proposed healthcare-related AI systems (Table~\ref{tab:evaluation}a). In contrast, \emph{Metric} refers to the quantitative measures used to compare LLM-as-a-Judge outputs with assessments from human evaluators (Table~\ref{tab:evaluation}b).

\begin{table*}[htbp]
\centering
\caption{Categorization of evaluation frameworks for LLM-as-a-Judge in healthcare research.}
\label{tab:evaluation}
\footnotesize

\noindent\textbf{(a) Rubric}~~{\color{gray}\small How is LLM-as-a-Judge used to evaluate novel healthcare data and frameworks?}\\[2pt]
\begin{tabularx}{\textwidth}{@{}p{1.5cm} p{5.3cm} p{3.9cm} p{3cm}}
\toprule
\textbf{Category} & \textbf{Definition} & \textbf{Rubrics} & \textbf{Example Studies} \\
\midrule
Behavior & Whether the judged output reflects appropriate reasoning behavior or action alignment. & memory rationality; action alignment &  \cite{song2026demma} \\
\addlinespace[2pt]
Communication & Whether the judged output communicates with appropriate empathy, naturalness, authenticity, or emotional fit. & empathy; personality consistency; language naturalness; authenticity; emotional reasonableness & \cite{song2026demma, yao2025biased, she2025emplifai, kafi2026reasoning, li2025counselbench} \\
\addlinespace[2pt]
Completeness & Whether the judged output covers the needed information without important omissions. & completeness; coverage; thoroughness & \cite{adib2026assessing, hisada2025filling, poore2025context, piya2026agenticsum} \\
\addlinespace[2pt]
Factuality & Whether the judged output is medically correct, faithful to evidence, and free from hallucinated claims. & faithfulness; correctness; hallucination; answer correctness; medical consistency & \cite{kocaman2025clinical, vasilev2025evaluating, adib2026assessing, remaki2026syncabel} \\
\addlinespace[2pt]
Presentation & Whether the judged output is concise, clear, readable, and logically organized. & coherence; succinctness; readability; organization; synthesis & \cite{poore2025context, she2025emplifai, piya2026agenticsum, shah2025tn} \\
\addlinespace[2pt]
Relevance & Whether the judged output is pertinent to the question, context, or clinical task. & answer relevancy & \cite{boll2025distillnote, croitoru2026privacy, poore2025context, zheng2025llm} \\
\addlinespace[2pt]
Safety & Whether the judged output avoids unsafe, risky, or ethically problematic content. & safety & \cite{zhuang2025towards, aali2025medval, she2025emplifai} \\
\addlinespace[2pt]
Utility & Whether the judged output is useful or helpful for the intended clinical purpose. & helpfulness; usefulness & \cite{liu2026tailored, wang2025healthq, sayeed2025rag, wu2025beyond} \\
\bottomrule
\end{tabularx}

\vspace{6pt}

\noindent\textbf{(b) Metric}~~{\color{gray}\small What metrics are used to assess the reliability of results generated by LLM-as-a-Judge?}\\[1pt]
\begin{tabularx}{\textwidth}{@{}p{1.5cm} p{4.5cm} p{4.8cm} p{3cm}}
\toprule
\textbf{Category} & \textbf{Definition} & \textbf{Metrics} & \textbf{Example Studies} \\
\midrule
Association & Continuous-score or rank association between judge outputs and reference scores. & Pearson correlation coefficient; rank correlation; Spearman's rank correlation coefficient; Kendall's tau; $R^2$ & \cite{vasilev2025evaluating, kocaman2025clinical, luo2025dialogguard, yao2026medqa} \\
\addlinespace[2pt]
Classification & Discrete-label performance against human labels, expert labels, or a ground truth. & accuracy; F1-score; recall; precision; specificity; sensitivity; AUC/AUROC; exact match; balanced accuracy; clinician-confirmed false negative rate & \cite{liu2025medq, niculae2025dr, jeong2026tool, peled2025dementia} \\
\addlinespace[2pt]
Efficiency & Runtime, time, or cost of judge evaluation. & runtime; cost & \cite{vasilev2025evaluating, zhou2025automating, williams2025human, fan2026halluhard} \\
\addlinespace[2pt]
Error & Continuous error or loss between judge outputs and reference values. & root mean square error; mean absolute error; mean squared error; Hamming loss & \cite{vasilev2025evaluating, badawi2026can} \\
\addlinespace[2pt]
Inference & Statistical inference for comparisons or uncertainty around metric estimates. & confidence interval; Wilcoxon signed-rank test; Friedman test; Cohen's $d$ & \cite{vasilev2025evaluating, williams2025human, bolpagni2025valise, kim2026pair} \\
\addlinespace[2pt]
Preference & Pairwise, head-to-head, win-rate, or preference-based comparisons. & win rate; preference rate; pairwise comparison; ties; LLM wins; vendor wins & \cite{hosseini2024benchmark, gunjal2025rubrics, wang2025chatthero, digiacomo2025guide} \\
\addlinespace[2pt]
Reference & Reference-based text similarity or retrieval-context measures. & ROUGE; BERT; BLEU; METEOR; context precision; context recall; context relevance & \cite{adib2026assessing, ferdousi2025rhealthtwin, wang2025healthq} \\
\addlinespace[2pt]
Reliability & Agreement or consistency between judges, humans, models, or repeated ratings. & agreement rate; Cohen's kappa; inter-rater agreement; Krippendorff's alpha; intraclass correlation coefficient; Gwet's AC1/AC2; Fleiss' kappa & \cite{li2026scaling, kocaman2025clinical, jarchow2025benchmarking, remaki2026syncabel} \\
\addlinespace[2pt]
Scale & Rubric, Likert, benchmark, scalar, or average score used as a measurement scale. & rubric score; Likert scale; HealthBench score; PDSQI-9 rubric score; CRIT score; mean/average score; scalar rating (0--5) & \cite{Cheng2026scaling, croxford2025evaluating, yao2026medqa, williams2025human} \\
\bottomrule
\end{tabularx}

\end{table*}

\textbf{Evaluation Rubrics.} Our analysis identifies eight rubric types used in collected studies (Table~\ref{tab:evaluation}a). A single study may use multiple rubric types. \emph{Behavior} refers to rubrics that use LLM-as-a-Judge to assess specific behaviors, reasoning processes, or action alignment~\cite{song2026demma}. For example, \citet{song2026demma} evaluate a dementia simulation agent by assessing memory rationality, including whether forgetting, repetition, and cue responses are consistent with the dementia profile, and action alignment, including whether nonverbal actions were plausible and consistent with verbal cues and clinical characteristics. \emph{Communication} refers to rubrics that evaluate communicative aspects of AI outputs~\cite{song2026demma, yao2025biased, she2025emplifai, kafi2026reasoning, li2025counselbench}. These aspects include empathy~\cite{song2026demma, yao2025biased, she2025emplifai}, language naturalness~\cite{song2026demma}, authenticity~\cite{song2026demma}, and emotional fit~\cite{song2026demma, yao2025biased, liu2026tailored}. Some studies further decompose communication quality into more specific dimensions. For example, in a dataset designed to support individuals in distress, \citet{liu2026tailored} used LLM-as-a-Judge to assess emotional supportiveness and cultural awareness.

Existing studies also evaluate \emph{Completeness}, which assesses whether all key information is included~\cite{adib2026assessing, hisada2025filling, poore2025context, piya2026agenticsum}, and \emph{Factuality}, which assesses whether the output is accurate and free from hallucinated claims~\cite{vasilev2025evaluating, curran2024examining, chen2025multiagent, kocaman2025clinical}. In addition, prior work has used LLM-as-a-Judge to evaluate \emph{Presentation}, including conciseness, clarity, and readability, and \emph{Relevance}, referring to the degree to which an output addresses the question, context, or clinical task. Several studies assess both presentation and relevance~\cite{kocaman2025clinical, madrid2025optimising, poore2025context, wang2025healthq, chen2025multi}. In addition, \emph{Safety} captures rubrics that use LLM-as-a-Judge to assess whether outputs contain unsafe, risky, or ethically problematic content. For example, \citet{adib2026assessing} evaluate safety in iCliniq medical QA tasks based on LLM judges, including the appropriateness of safety disclaimers, avoidance of harmful advice, and recommendations to consult healthcare professionals when appropriate, using a 1--5 rating scale. Finally, \emph{Utility} refers to rubrics that assess whether an output is useful or helpful for the intended clinical purpose~\cite{liu2026tailored, wang2025healthq, sayeed2025rag, wu2025beyond}.

\textbf{Evaluation Metrics.} To understand how existing studies assesses the reliability of LLM-as-a-Judge generated inferences, we also identify multiple categories of metrics used in existing work (Table~\ref{tab:evaluation}b). Our analysis focuses on three major healthcare application domains (as identified in Section~\ref{sec:health_applications}). While these domains share some common evaluation paradigms, each exhibits distinct emphases reflecting its task characteristics.

\textbf{\textit{Metrics for Clinical Decision Support}}.
Across \textit{Clinical Decision Support} studies, LLM-as-a-Judge is evaluated with a diverse but coherent set of metrics. One evaluation assesses alignment with human experts, commonly using agreement measures such as Cohen's $\kappa$, Krippendorff's $\alpha$, Fleiss' $\kappa$, or agreement rate, together with correlation-based metrics such as Pearson's $r$, Spearman's $\rho$, and Kendall's $\tau$ to measure concordance with clinician ratings~\cite{wu2025automated}. When LLM judges produce discrete labels, studies often report classification metrics, including accuracy, precision, recall, F1-score, and ROC-AUC~\cite{sarvari2025rapidly}; when judges produce continuous scores, error-based metrics such as MAE, MSE, and RMSE are also used~\cite{badawi2026can}. Many studies further apply rubric-based or Likert-scale evaluations across clinically relevant dimensions, as in LiveMedBench~\cite{yan2026livemedbench} and CounselBench~\cite{li2025counselbench}. Beyond predictive performance, several studies examine judge reliability and robustness through inter-judge agreement, consistency, and task-specific downstream metrics such as diagnostic recall or safety detection~\cite{stamatis2026beyond}.

\textbf{\textit{Metrics for Clinical NLP}}.
Evaluation in \textit{Clinical NLP} places particular emphasis on fine-grained textual quality, factual grounding, and workflow-level utility. Many studies use multi-dimensional rubric or Likert-scale evaluations to assess factual accuracy, hallucination, completeness, coherence, conciseness, and clinical actionability, reflecting the practical requirements of generated notes, summaries, and extracted information~\cite{piya2026agenticsum, aali2025medval, croxford2025evaluating}. A distinctive feature of this domain is claim-level factuality verification, in which outputs are decomposed into atomic statements and judged as supported, contradicted, or missing relative to source data~\cite{chung2025verifact}. \textit{Clinical NLP} studies also often embed LLM judges within end-to-end systems for summarization, information extraction, or de-identification, where task-specific downstream metrics (e.g., hallucination detection rate, extraction accuracy, and de-identification precision and recall) are used to evaluate operational performance~\cite{lilli2026prompt,miranda2025mamma}. In addition, prior research assesses robustness and reproducibility through repeated runs and cross-judge consistency~\cite{chen2025multiagent}. Overall, compared with \textit{Clinical Decision Support}, metrics in \textit{Clinical NLP} are more text-centric and decomposition-driven, with emphasis on fine-grained rubric design and assessment of document-level clinical workflows.

\textbf{\textit{Metrics for Medical Knowledge \& QA}}.
In this domain, LLM-as-a-Judge evaluation places greater emphasis on reasoning quality, factual correctness, and answer justification, reflecting the explanatory nature of medical QA tasks. A key feature is the use of reasoning-oriented rubrics, in which judges assess logical validity, clinical reasoning quality, evidence use, and safety. For example, \citet{zhou2025automating} use Likert-style reasoning scores aligned with expert ratings, while \citet{chen2025gaps} adopt multi-dimensional rubric frameworks to capture more nuanced reasoning performance. Another prominent feature is the use of pairwise or ranking-based metrics (e.g., win rate, rank agreement, and Kendall's $\tau$) are often used to assess whether LLM judges can distinguish higher-quality answers from weaker alternatives~\cite{de2025ranking}.

Medical QA evaluation also emphasizes fine-grained factuality and statement-level error analysis. Several studies quantify correctness using counts or proportions of correct, incorrect, or missing facts, or through structured factuality scores~\cite{steinigen2026fact}. This granular approach enables targeted assessment of hallucination and misinformation beyond a single overall score. The domain also incorporates robustness- and bias-oriented evaluation, such as sensitivity to adversarial inputs, misleading context, and cultural cues~\cite{liu2024decoy, rezaei2026counterfactual}. Several studies further include meta-evaluation and efficiency measures, such as judge--expert correlation, statistical discrimination tests, evaluation cost, runtime, and confidence calibration~\cite{zhou2025automating,anantha2025nanoflux}. Compared with other application domains, medical QA evaluation is more reasoning-centric and comparison-driven, with emphasis on ranking consistency, fine-grained factual verification, and robustness to adversarial or contextual variation.

\subsection{LLM-as-a-Judge Alignment with Human Annotators}
\label{sec:judge_alignment}

Because LLM judges are imperfect evaluators, their reliability should be assessed by the extent to which their judgments align with human annotations. Human-LLM alignment is therefore central to validating LLM-as-a-Judge methods in healthcare. However, studies quantify this alignment using different metrics depending on task structure, output format, and rating scale. Commonly reported metrics include raw agreement rate, Cohen's $\kappa$, Pearson's $r$, accuracy, F1-score, Spearman's $\rho$, win rate, Kendall's $\tau$, and Krippendorff's $\alpha$ (see Section~\ref{sec:judge_evaluation} for specific explanations). To support a comparable synthesis, we focus on three widely reported metrics, including agreement rate, Cohen's $\kappa$, and correlation, and use them to illustrate the reliability of LLM-as-a-Judge methods across healthcare tasks in the current literature.

\begin{figure*}[ht]
    \centering
    \includegraphics[width=1.0\linewidth]{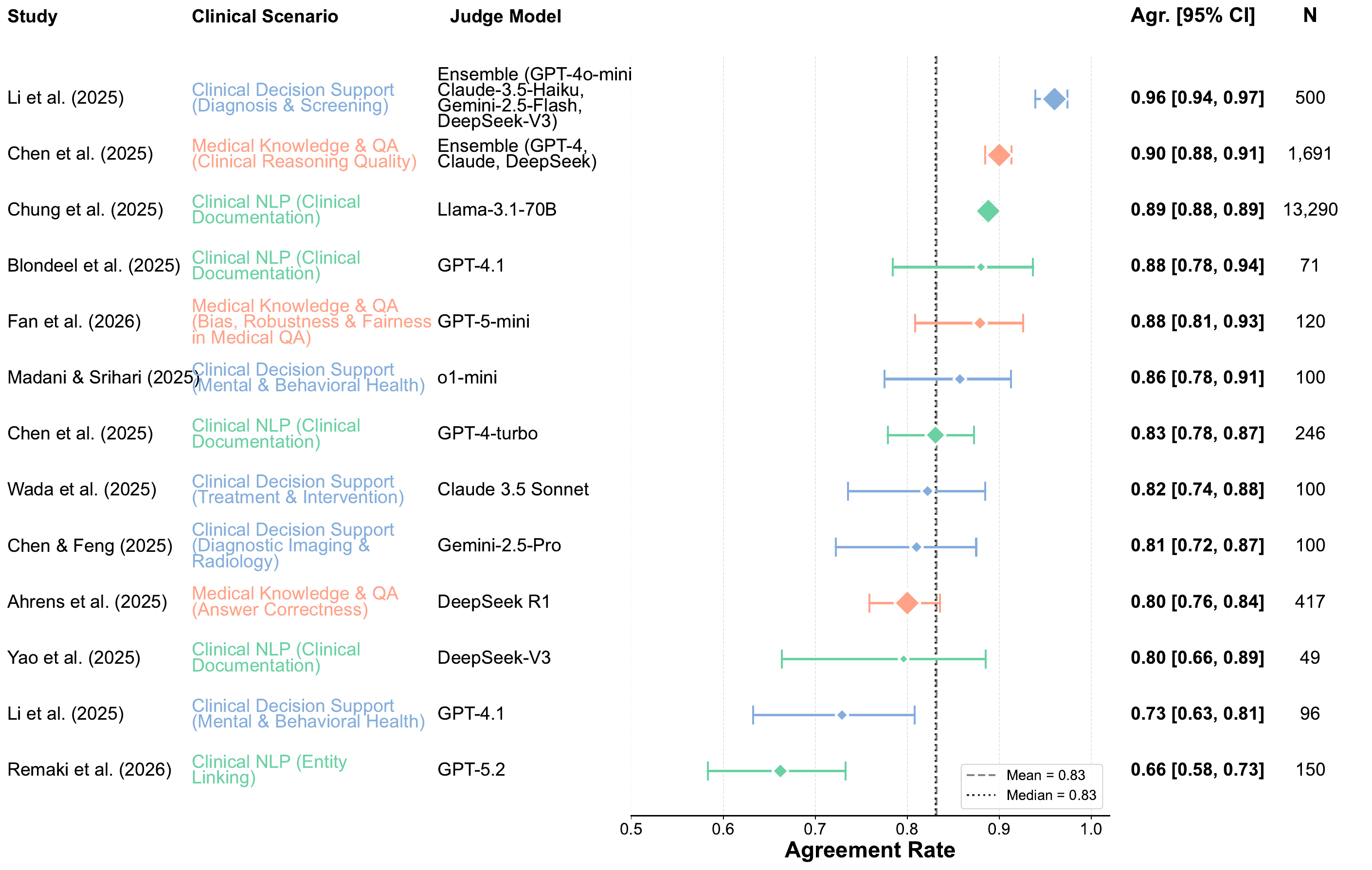}
    \caption{\textbf{Agreement rate between LLM-as-a-Judge and human experts} Forest plot of 13 examined studies reporting agreement rate with 95\% confidence intervals sorted by agreement rate. The dotted line marks the median agreement rate (0.83), and the dashed line marks the mean agreement rate (0.83) across the 13 examined studies. Agreement values are rounded to two decimal places.}
    \label{fig:meta_agreement}
\end{figure*}

\textbf{Agreement rate with human experts.}
Across 13 examined studies reporting agreement against expert judgments (Figure~\ref{fig:meta_agreement}), judge-expert concordance ranges from 0.66 to 0.96 (median 0.83, mean 0.83). The lowest value is observed in fine-grained biomedical entity linking, where \textit{GPT-5.2} reaches 0.66 in distinguishing Correct, Broad, Narrow, and No-relation labels~\cite{remaki2026syncabel}; the highest value appears in MedGUIDE, where an ensemble of \textit{GPT-4o-mini}, \textit{Claude-3.5-Haiku}, \textit{Gemini-2.5-Flash}, and \textit{DeepSeek-V3} is used to score guideline-grounded multiple-choice questions, with validation on 500 human-reviewed samples~\cite{li2025medguide}. In addition, we observe that LLM-as-a-Judge ensembles cluster near the top, including a three-judge ensemble (\textit{GPT-4}, \textit{Claude}, \textit{DeepSeek}) that reaches 0.90 on rubric-anchored adequacy and safety items in NSCLC care~\cite{chen2025gaps}, supporting the view that aggregating heterogeneous judges can reduce idiosyncratic bias. Well-scoped tasks with anchored rubrics generally outperform open-ended generation: structured factuality checks on Brief Hospital Course narratives reach 0.89 across 13,290 propositions~\cite{chung2025verifact}, whereas mental-health counseling rubrics with more subjective language register 0.73~\cite{li2025counselbench}. Finally, raw agreement can be sensitive to the reference standard: in one study, the same judge that reached 0.83 agreement with at least one expert dropped to 0.51 against majority-of-experts judgments~\cite{chen2025multiagent}, highlighting how expert disagreement and subjective criteria can affect apparent judge reliability.

\begin{figure*}[ht]
    \centering
    \includegraphics[width=1.0\linewidth]{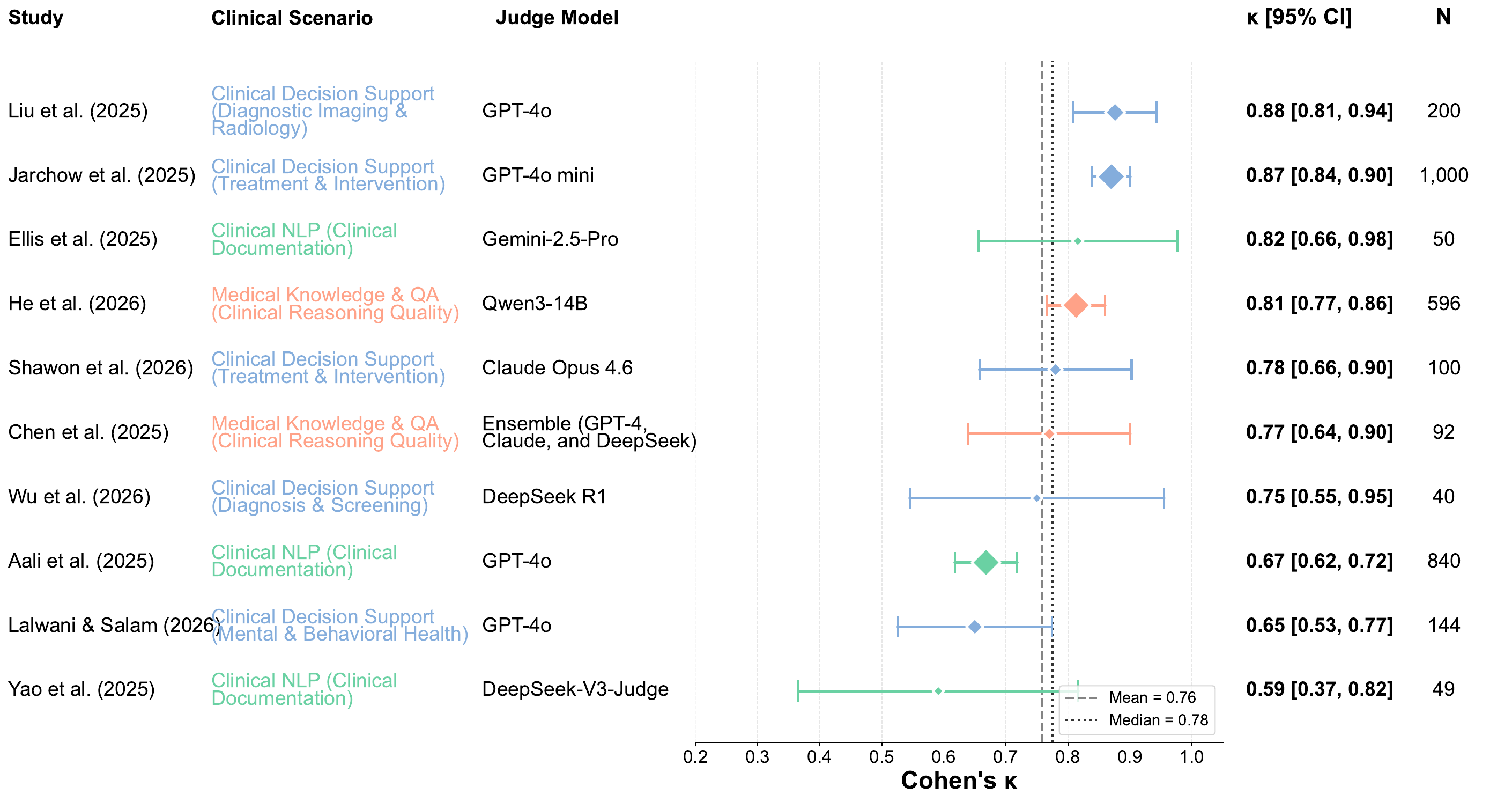}
    \caption{\textbf{Chance-corrected agreement (Cohen's $\kappa$) between LLM-as-a-Judge and human experts.} Forest plot of 10 studies reporting Cohen's $\kappa$ with 95\% confidence intervals and validation sample sizes ($N$), sorted by $\kappa$. The dotted line marks the median Cohen's $\kappa$ (0.78), and the dashed line marks the mean Cohen's $\kappa$ (0.76) across the 10 examined studies. Cohen's $\kappa$ values are rounded to two decimal places.}
    \label{fig:meta_kappa}
\end{figure*}

\textbf{Chance-corrected agreement (Cohen's $\kappa$).}
Across 10 examined studies reporting Cohen's $\kappa$ (Figure~\ref{fig:meta_kappa}), values range from 0.59 to 0.88, broadly covering the ``moderate'' to ``strong'' agreement range under conventional benchmarks. The lowest value ($\kappa = 0.59$) is observed for DeepSeek-V3 evaluating educator dialogues and personalized discharge summaries~\cite{yao2025dischargesim}. The highest value is reported for \textit{GPT-4o} in medical image quality assessment ($\kappa = 0.88$)~\cite{liu2025medq}, followed by \textit{GPT-4o-mini} for personalized longevity recommendations ($\kappa = 0.87$)~\cite{jarchow2025benchmarking}. \textit{Claude-Opus-4.6} also shows good alignment with human annotators for unperturbed medical concept validation ($\kappa = 0.78$), but its performance drops substantially for perturbed concepts ($\kappa = 0.24$)~\cite{shawon2026advancing}. More subjective or multi-dimensional rubric tasks tend to show lower reliability; for example, four-class clinical risk grading in MedVAL remains in the moderate range ($\kappa = 0.67$)~\cite{aali2025medval}. Agreement is also task-dependent within the same judge family: \textit{GPT-4o} reaches 0.88 in medical image quality assessment~\cite{liu2025medq}, but shows lower agreement on mental and behavioral health prompts~\cite{lalwani2026supportiveness}. Finally, the gap between raw agreement and $\kappa$ is informative. \citet{yao2025dischargesim} report 0.80 raw agreement but $\kappa = 0.59$ on the same dataset, showing that prevalence-corrected reliability can be substantially lower when one rating category dominates.

\begin{figure*}[ht]
    \centering
    \includegraphics[width=1.0\linewidth]{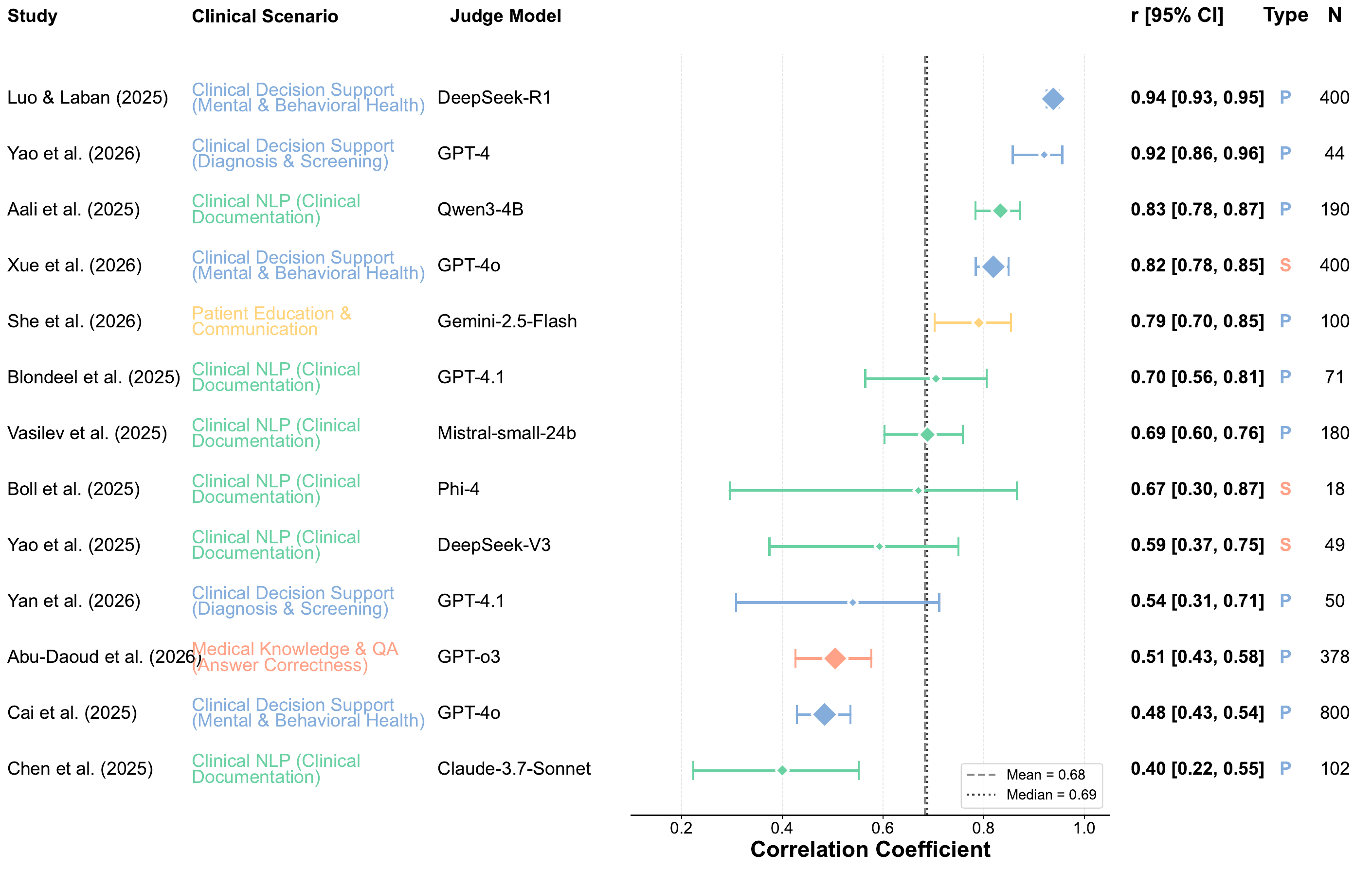}
    \caption{\textbf{Score-level correlation between LLM-as-a-Judge and human expert ratings.} Forest plot of 13 studies reporting Pearson's $r$ (P, $n=10$) or Spearman's $\rho$ (S, $n=3$) with 95\% confidence intervals, sorted by correlation. The dotted line marks the median correlation value (0.69), and the dashed line marks the mean correlation value (0.68) across the 13 examined studies. Correlation values are rounded to two decimal places.}
    \label{fig:meta_correlation}
\end{figure*}

\textbf{Score-level correlation with experts.} Across 13 examined studies reporting Pearson or Spearman correlation (Figure~\ref{fig:meta_correlation}; 10 Pearson, 3 Spearman), judge--expert correlation ranges from 0.40 to 0.94 (median 0.69, mean 0.68). \textit{DeepSeek-R1} reaches $r = 0.938$ on the behavior dimension of a five-dimension psychosocial safety rubric~\cite{luo2025dialogguard}, \textit{GPT-4} attains $r = 0.92$ on OSCE-style InfoGatherQA~\cite{yao2026medqa}, and MedVAL-fine-tuned \textit{Qwen3-4B} reaches $r = 0.833$ on a clinical-summary subset~\cite{aali2025medval}. In contrast, \textit{GPT-4o} reaches only $r = 0.483$ on counseling responses evaluated across five subjective safety dimensions~\cite{cai2025exploring}. The lower end of the distribution is concentrated in open-ended counseling and multi-dimensional clinical text evaluation, whereas the upper end includes tasks like psychosocial safety, clinical-skills scoring, and clinical summary. In addition, it should be highlighted that score-level correlation and agreement capture different aspects of judge reliability. Correlation indicates whether LLM judges preserve the relative ordering of expert scores, whereas agreement metrics assess consistency in assigned labels or categories. These metrics therefore capture different forms of human--LLM alignment and should be interpreted separately when comparing judge reliability as seen in prior studies~\cite{li2025counselbench, han2026optimizing, jang2025medtutor}.

\section{DISCUSSION}
\label{sec:discussion}

The rapid growth of LLM-as-a-Judge research in healthcare (Figure~\ref{fig:trend}) suggests a broad change in how clinical AI systems are evaluated. As these systems increasingly produce open-ended outputs---diagnostic narratives, counseling responses, discharge summaries, and patient-facing explanations---the evaluation gap of fixed-answer benchmarks and conventional metrics for unstructured clinical data have become more pronounced. LLM-as-a-Judge has emerged as a practical response to this gap, offering a way to assess content along dimensions (e.g., reasoning quality, factual consistency, completeness, safety, empathy, and clinical usefulness) that are difficult to formalize~\cite{gu2024survey, croxford2025evaluating, liu2025medq, xue2026privacypreserving, ding2025medbench, li2025counselbench, chen2025multi}. In response to \textbf{RQ4}, we draw on the reviewed studies to identify the opportunities and challenges that LLM-as-a-Judge holds for future healthcare AI evaluation. We then discuss considerations for deploying LLM-as-a-Judge, along with the limitations of this review.

\subsection{Potential of LLM-as-a-Judge in Healthcare}

\textbf{Scaling evaluation across unstructured clinical content.} Much of clinical practice is recorded in unstructured text, where lexical-overlap metrics and fixed-answer benchmarks often miss clinical correctness or contextual appropriateness~\cite{adnan2020role,malmasi2018extracting}. Expert annotation can capture the clinical nuance but is costly, slow, and difficult to apply at the volumes required for modern AI development and monitoring~\cite{malmasi2018extracting,tayefi2021challenges}. LLM-as-a-Judge offers a path through this trade-off by approximating rubric-based expert judgment on free-text outputs at substantially reduced time and cost. The popular application areas in this review, including \textit{Clinical Decision Support}, \textit{Clinical NLP}, \textit{Medical Knowledge \& QA}, and \textit{Medical Communication}, are precisely the settings where this bottleneck is most evident, as each generates open-ended outputs at volumes that exceed the capacity of expert review. The potential of LLM-as-a-Judge is therefore not to replace clinician review, but to enable evaluation at a scale supporting routine monitoring, iterative refinement, and continuous improvement of clinical AI systems.

\textbf{Moving from correctness to clinically meaningful endpoint.} The reviewed studies show that LLM-as-a-Judge is often used for more than checking whether an answer is correct. Many rubrics evaluate whether an output is factual, complete, safe, readable, and clinically useful (Table~\ref{tab:evaluation}). This broader evaluation is important because clinical errors are not always captured by accuracy alone. For example, prior work has shown that fluent clinical summaries can contain hallucinated facts or omit critical relevant information~\cite{vasilev2025evaluating, saito2025generation}. Similarly, long-form medical Q\&A benchmarks suggest that correctness alone is insufficient for assessing clinical answer quality~\cite{hosseini2024benchmark}. This suggests that LLM-as-a-Judge is informative when the rubric is tied to the specific clinical output, such as PDSQI-9 for documentation quality~\cite{croxford2025evaluating}, OSCE-style criteria for clinical skills~\cite{yao2026medqa}, or atomic-claim factuality assessment against EHR evidence~\cite{chung2025verifact}. By contrast, generic prompts that ask for overall ``quality'' are less likely to produce clinically interpretable judgments.

\textbf{Matching judging strategy to clinical risk.} A third potential lies in the growing ability to calibrate LLM-as-a-Judge architecture to the risk profile of the clinical task, rather than applying a uniform evaluator across heterogeneous settings. Although prompt engineering appears in nearly all studies, the technical toolkit has expanded to include ensembles, multi-agent designs, retrieval grounding, and fine-tuning, giving researchers concrete options for tuning evaluation depth to clinical stakes. A single rubric-based judge may suffice for structured factuality checks or low-risk documentation review. For safety-sensitive tasks such as psychosocial risk assessment or treatment recommendation, multi-agent or persona-based designs can expose disagreement that would be hidden in a single score~\cite{luo2025dialogguard, chen2025multiagent}. For knowledge-intensive judgments, retrieval can supply guidelines, drug information, or patient records directly to the judge~\cite{sarvari2025rapidly, yan2026livemedbench}. This flexibility opens the possibility of risk-stratified evaluation pipelines, in which lightweight judges handle routine monitoring while more advanced multi-agent or retrieval-grounded designs are reserved for high-stakes clinical content.

\textbf{Building toward auditable evaluation co-pilots.} Another potential concerns how LLM judges can be integrated with expert workflows. Across examined studies reporting quantitative validation (Section~\ref{sec:judge_alignment}), the evidence is encouraging. For example, agreement rate ranges between 0.66 and 0.96 (Figure~\ref{fig:meta_agreement}), while median Cohen's $\kappa$ is between 0.59 and 0.93 (Figure~\ref{fig:meta_kappa}). These results indicate that strong alignment with experts is achievable, particularly when the task is well scoped and the rubric is explicit. They also point to a potential path forward. Rather than positioning LLM judges as substitutes for clinical expertise, the more promising role is as evaluation co-pilots that handle routine assessment at scale while routing uncertain or high-stakes cases to human experts. This could expand the scale of clinical AI evaluation without diminishing expert oversight, with human experts increasingly focused on the cases where their judgment adds the most value.

\textbf{Bridging the methodological-to-clinical translation gap.} A final potential lies in the translational deployment of LLM-as-a-Judge systems. The methodology originated in the computer science community, and publications continue to be disseminated primarily as preprints (46 arXiv studies among our 134 filtered records). While this partly reflects the fast pace of AI methodology iteration, the limited representation in PubMed-indexed journals signals unrealized translational opportunities: fields such as medical informatics, decision support, mental and behavioral health, patient communication, and medical education stand to benefit substantially from the scalable evaluation pipelines enabled by LLM-as-a-Judge, but peer-reviewed publications in relevant venues are still sparse. This is particularly notable given the moderate-to-high performance of such systems observed across these fields in our review. We anticipate that LLM-as-a-Judge is approaching a critical point at which deployment can deliver meaningful benefit to healthcare practitioners and patients, particularly in the high-performance subdomains synthesized here.

\subsection{Failure Modes of LLM-as-a-Judge in Healthcare}

\textbf{Evaluation bias from shared model families.} Bias arising from judges evaluating outputs generated by models in the same family has been examined in prior work~\cite{li2026scoping}. Specifically, when both the generator and the evaluator belong to the same model family, they often share similar training data distributions, inductive biases, and knowledge gaps. As a result, a GPT-based judge evaluating GPT-generated clinical text may assign high scores to outputs containing errors that GPT-family models systematically fail to recognize, not because such errors are absent, but because both systems lack the capacity to identify them reliably. This creates a form of correlated evaluation bias in which shared blind spots can artificially inflate perceived performance.

\textbf{Conflating surface presentation with substantive quality.} LLM judges often struggle to consistently interpret and apply evaluation criteria, conflating surface-level features with substantive quality. For example, linguistic fluency may be rewarded despite factual inaccuracies, clearer organization mistaken for greater completeness, and an assertive tone interpreted as a marker of professional credibility~\cite{vasilev2025evaluating, williams2025human, de2025ranking}. In clinical document summarization, LLM judges perform poorly on redundancy, coherence, hallucination detection, and grammar, and in some cases correlate less strongly with expert assessments than conventional metrics such as BERTScore~\cite{vasilev2025evaluating}. Similar findings appear in medical education, where ``ExaminerGPT'' grades leniently until explicitly instructed to mark more strictly~\cite{saggar2026ai}. These findings suggest that LLM judges do not reliably execute rubrics as formal decision rules, but instead simulate what scoring behavior should look like based on textual cues.

\textbf{Insufficient depth in clinical semantic reasoning.} A related failure mode concerns shallow domain reasoning beneath fluent biomedical language. Although frontier models encode broad biomedical knowledge, they continue to struggle when evaluating nuanced clinical outputs. In ICD-10-CM prediction studies, LLM judges misclassified chronic versus newly inferred diagnoses, mishandled historical conditions, and misinterpreted prompt terminology, leading to inflated estimates of model performance~\cite{dai2025model}. In biomedical relation extraction, even domain-adapted judges with structured outputs remained limited in recognizing complex relations, ambiguous terminology, and implicit entity connections~\cite{laskar2025improving}. These findings indicate that fluency in medical language should not be conflated with competence in clinical semantics, a distinction that becomes consequential when judges are deployed on tasks requiring fine-grained clinical reasoning.

\textbf{Evaluation hallucination.} Prior work on hallucination has focused primarily on generation models inventing facts, but several studies show that evaluator models can hallucinate as well. LLM judges may misdescribe candidate responses, fabricate flaws, or even alter the task definition itself. In one trustworthiness analysis, LaaJ-alpha (a \textit{GPT-4o}-based prototype) is found to ``solve'' matching tasks by changing the underlying matching problems rather than correctly evaluating the original task specification~\cite{curran2024examining}. In clinical document summarization, judges are similarly unreliable in detecting factual hallucinations in generated outputs~\cite{vasilev2025evaluating}. This poses a particular safety concern: when both the generator and the evaluator are LLMs, their errors can reinforce rather than correct one another. Automated evaluation, in this case, no longer provides an independent check on generation quality.

\textbf{Prompt sensitivity and cross-linguistic fragility.} Robustness of LLM judges remains contingent on prompt design and language environment. Minor changes in prompt wording can alter grading strictness, score distributions, and rationale quality~\cite{saggar2026ai}. In global health evaluations, performance deteriorates and costs increases when moving from English to Kinyarwanda~\cite{williams2025human}. These findings indicate that current LLM judges remain sensitive to prompt design and language resources, raising concerns for both reproducibility and equity across clinical settings.

\subsection{Considerations of LLM-as-a-Judge Deployment in Clinical Settings}

Moving LLM-as-a-Judge from retrospective evaluation to real-world clinical deployment fundamentally changes its role: the judge is no longer merely a measurement tool, but an active component of the clinical workflow that may directly influence downstream decisions and patient care. In practice, lower-risk applications such as documentation screening, summary review, and preliminary factuality checking may represent more feasible early deployment settings, where LLM judges can function as triage systems that identify potentially problematic cases for clinician review rather than fully autonomous evaluators. However, broader clinical deployment still requires careful consideration of safety, reliability, and human oversight.

\textbf{Clinical risks remain a major barrier to deploying LLM judges in healthcare.} Errors in judging diagnostic reasoning or treatment recommendations may propagate unsafe clinical decisions \cite{vasilev2025evaluating, williams2025human, ding2025medbench}, while inaccuracies in evaluating clinical notes or discharge summaries may compromise communication, billing, and medico-legal records \cite{chen2025multiagent, shah2025tn}. Importantly, overly favorable or overconfident judgments may create false reassurance and obscure unsafe model outputs, particularly when healthcare LLMs generate fluent but factually incorrect or clinically inappropriate responses \cite{vasilev2025evaluating, shah2025tn, ding2025medbench}. In addition, disagreement across judges and low-confidence assessments may serve as useful uncertainty signals that trigger additional human review \cite{liu2025statistically, badawi2026can}. Even when LLM judges demonstrate strong average performance, limited clinician trust in automated judge systems may still necessitate continued manual verification.

\textbf{Healthcare heterogeneity limits model generalizability.} Healthcare settings vary widely in patient populations, disease prevalence, EHR systems, documentation practices, clinical workflows, local policies, resources, and language use. As a result, a judge that performs well on a benchmark or within one hospital system may not perform equally well in another setting~\cite{dai2025model,williams2025human,aali2025medval}. This limits the portability and external validity of LLM-as-a-Judge frameworks. A further concern is that LLM judges may inherit demographic, linguistic, and socioeconomic biases from their underlying foundation models. These biases could lead to uneven evaluation quality across patient subgroups, institutions, or care environments~\cite{williams2025human, hisada2025filling}.

\textbf{Continuous monitoring is necessary for reliable deployment.} Clinical standards change over time as new guidelines, therapies, evidence, and local workflows emerge. Without regular updates or re-validation, an LLM judge may continue to apply outdated criteria or produce evaluations that no longer align with current standards of care~\cite{yan2026livemedbench,ding2025medbench}. Reliable deployment also requires transparent reporting of the judge model, prompt, validation data, calibration behavior, measures, and common error patterns~\cite{liu2025statistically,sarvari2025rapidly}. In safety-critical settings, false-negative judgments are especially important because they may allow unsafe outputs to pass without human review~\cite{chen2025multiagent,vasilev2025evaluating}. For this reason, deployed LLM-as-a-Judge systems should include audit logs, version tracking, drift monitoring, and periodic re-validation. These governance mechanisms can help preserve interpretability, accountability, and reliability as models and clinical workflows evolve~\cite{yadav2025sees,li2026scaling}.

\textbf{Patient-facing use raises ethical and legal concerns.} These concerns are especially important when LLM judges are used to evaluate patient--LLM conversations or patient education materials. In these settings, automated judgments may shape how clinical information is communicated to patients, raising questions about liability, informed consent, privacy, transparency, and accountability~\cite{bentley2026vera,liu2026tailored}. The risks are amplified when judging workflows involve protected health information. Such workflows require secure data handling, clear audit trails, and compliance with relevant privacy and regulations during both model development and deployment~\cite{aali2025medval,thomas2025preserving,wu2025chain}.

\textbf{Prospective evaluation in the intended care workflow.} An important consideration for clinical deployment of LLM-as-a-Judge is evidence from prospective evaluation in the intended care workflow. Most evidence for LLM-as-a-Judge is currently retrospective, relying on archived cases, benchmark datasets, or post-hoc comparison with expert annotations. While essential for system development and initial validation, retrospective studies may not capture real-world workflow constraints, clinician interaction, alert fatigue, automation bias, latency, or the safety implications of hallucinated outputs. Prospective ``silent mode" evaluation, in which LLM-as-a-Judge outputs are generated for real clinical cases but not shown to clinicians or used to alter care, may provide a useful intermediate step for assessing reliability and failure modes before broader clinical deployment~\cite{tikhomirov2026scoping, kwong2022silent}. For applications in which LLM-as-a-Judge influences clinical decision-making, pragmatic trials or other prospective implementation studies may be needed to assess clinical effectiveness, safety, and impact on the clinical workflow or clinician behavior~\cite{han2024randomised}.

\subsection{Limitations and Future Work}

This review has several limitations. First, limitations arise from evidence synthesis. Reported metrics (e.g., agreement, correlation, and pairwise preference) capture different aspects of judge performance under different task conditions. For this reason, our meta-analysis emphasizes cross-study patterns rather than deriving a single pooled estimate of judge performance across healthcare tasks. In addition, our extracted study-level variables and manually coded labels may be affected by annotation ambiguity, especially when papers provide incomplete descriptions of rubrics, judge prompts, validation samples, or human reference standards.

Another limitation comes from the design and reporting of the included studies. They differ in clinical task, judge model, rubric design, and validation protocol. As a result, the observed variation in judge performance may reflect differences in study design and reporting quality, rather than differences in the capability of LLM judges. Many studies illustrate the alignment with human experts but provide limited information on operational dimensions of LLM-as-a-Judge deployment, such as inference cost, latency, reproducibility, privacy safeguards, and integration with existing clinical workflows. These factors are likely to be decisive for real-world adoption but remain underreported in the current analysis.

Next, limitations stem from the composition and timing of the study corpus. Although we use a structured search strategy across multiple academic databases, LLM-as-a-Judge research is evolving rapidly, and many studies appear first as preprints or benchmark releases before they are indexed in major databases. This is reflected in the substantial proportion of preprints in our collection. The pace of LLM release further compounds this limitation, as our review may not fully capture the performance of the most recent frontier models, such as \textit{GPT-5.5}. In addition, the included studies are drawn predominantly from English-language settings, with limited representation of low-resource languages and underrepresented patient populations. Future work could continuously track new models and broaden geographic and linguistic coverage as the field evolves.

A further limitation concerns the actionable guidance this review can offer. We describe a broad set of methods, including prompt engineering, ensembles, multi-agent designs, RAG, fine-tuning, distillation, and calibration, together with diverse rubrics and metrics, but do not recommend configurations for specific clinical tasks. This restraint reflects the current state of the evidence: included studies vary along clinical task, judge model, rubric design, validation sample, and reference standard, making it difficult to isolate the contribution of any single design choice. To our knowledge, no large-scale controlled comparison has systematically varied judge architecture (e.g., single judge vs. ensemble vs. multi-agent), prompting strategy (e.g., rubric-only vs. CoT vs. few-shot), or grounding source (e.g., closed-book vs. retrieval-augmented) while holding the clinical task fixed. Future work could address this gap through benchmarking on shared healthcare evaluation corpora spanning multiple clinical scenarios, moving the field from a catalog of techniques toward evidence-based guidance on which judging strategies suit which clinical tasks.
\section{CONCLUSION}
\label{sec:conclusion}

This review demonstrates that LLM-as-a-Judge has emerged as a feasible approach for scalable evaluation in healthcare AI, with concentrated application in clinical decision support, clinical NLP, medical knowledge and QA, and medical communication. Across the 134 included studies, LLM judges are most commonly implemented through prompt-based evaluation, augmented by strategies such as ensembles, multi-agent designs, RAG, and fine-tuning. Evidence from human validation indicates that LLM judges can approximate expert judgments with moderate to strong alignment in many settings, although reliability varies substantially by task, rubric design, and model choice. LLM judges should therefore be understood not as replacements for expert evaluation, but as scalable tools that complement clinical expertise and require transparent, rigorous validation. Going forward, the field should establish when these judges are trustworthy, for which clinical tasks, and under what validation standards, so that LLM-as-a-Judge can be deployed as an auditable and clinically grounded component of healthcare AI evaluation.

\bibliographystyle{ACM-Reference-Format}
\bibliography{main}

\appendix

\section{Codebook for Data Extraction}
\label{app:codebook}

To support transparent and reproducible synthesis, we develop a structured codebook for extracting information from the included studies. The codebook contains fields organized into four groups: bibliographic metadata, study and clinical context, judge configuration, and evaluation and validation (Table~\ref{tab:codebook}). Each field is coded as a controlled vocabulary (CV), free text (FT), numeric (Num), or binary (Bin) entry, depending on whether a closed taxonomy or open description is appropriate. Controlled vocabularies are iteratively refined through pilot coding, with disagreements resolved by team discussion. Free-text fields preserve study-specific details that resist standardization, while numeric and binary fields support quantitative synthesis and inclusion checks for the meta analytics.

\begin{table}[ht]
\centering
\footnotesize
\caption{Data extraction codebook. The 18 fields used to code each of the 134 included studies, organized into four groups. ``Coding'' indicates whether the field was coded with a controlled vocabulary (CV), free text (FT), numeric (Num), or binary (Bin).}
\label{tab:codebook}
    \begin{tabular}{p{3cm} p{1cm} p{8.8cm}}
    \toprule
    \textbf{Field} & \textbf{Coding} & \textbf{Description and example values} \\
    \midrule
    \multicolumn{3}{l}{\textit{Group 1. Bibliographic metadata}} \\
    \midrule
    Paper title & FT & Title of the publication as it appears in the source record. \\
    Author & FT & Full author list, in publication order. \\
    Author affiliations & FT & Primary institutional affiliations of the authors, including department and country. \\
    Year-month & Num & Publication year and month in YYYY-MM format; preprint posting date used for non-peer-reviewed records. \\
    DOI & FT & Digital Object Identifier; arXiv ID recorded when no DOI was assigned. \\
    \midrule
    \multicolumn{3}{l}{\textit{Group 2. Study and clinical context}} \\
    \midrule
    Study dataset & FT & Underlying data source(s); e.g., MIMIC-IV, MedQA, institutional EHR cohort, simulated patient cases, curated clinician-authored question bank. \\
    Study data modality & CV & Input modality used by the system being judged: text only, text + EHR, text + radiology images, radiology images only, audio, or other multimodal. \\
    Clinical scenario & CV & Top-level clinical domain: clinical decision support, clinical NLP, medical knowledge \& QA, medical education, or other. \\
    Clinical task description & FT & Short free-text summary of the specific application; e.g., ``evaluating concordance between AI-generated and specialist eConsult responses,'' ``fact-checking long-form clinical narratives against patient EHR notes.'' \\
    \midrule
    \multicolumn{3}{l}{\textit{Group 3. Judge configuration}} \\
    \midrule
    Judge content & CV & Type of artifact being evaluated; e.g., AI content [EHR summaries], AI content [diagnosis], AI content [counseling response], biomedical data [medical question], biomedical data [image]. \\
    Judge content description & FT & Specific evaluation dimensions targeted by the rubric; e.g., factual consistency, hallucination, completeness, safety, reasoning quality, empathy. \\
    LLM judge model & FT & Model name, version, and provider for every model used as a judge; e.g., GPT-4o, Claude 3.5 Sonnet, Gemini 2.5 Pro, DeepSeek-V3, Qwen3-14B, Llama-3.1-70B. \\
    LLM technique category & CV & One or more of seven non-mutually-exclusive labels: Prompt Engineering, Ensemble, RAG, Fine-tuning, Multi-agent, Calibration, Distillation. \\
    LLM technique description & FT & Specific implementation; e.g., few-shot rubric prompting with chain-of-thought reasoning; self-supervised distillation from GPT-4o with QLoRA fine-tuning; majority voting over three heterogeneous judges. \\
    \midrule
    \multicolumn{3}{l}{\textit{Group 4. Evaluation and validation}} \\
    \midrule
    Judge evaluation metrics & FT & Full list of statistics reported in the original study; e.g., percent agreement, Cohen's $\kappa$, Krippendorff's $\alpha$, Pearson's $r$, Spearman's $\rho$, F1, ROC-AUC, win rate. \\
    Judge performance & FT & Corresponding numerical results, including point estimates and 95\% confidence intervals where reported. \\
    Judge sample size & Num & Number of items in the validation set used to compute judge--human agreement; sub-sampling design noted where applicable. \\
    Judge against human validation & Bin & Whether the judge was directly compared against expert human ratings (Yes/No); precondition for inclusion in the meta-analytic figures. \\
    \bottomrule
\end{tabular}
\end{table}

\section{Geographic Distribution and Institutional Collaboration}
\label{app:geo_collaboration}

\begin{figure*}[ht]
\centering
\includegraphics[width=0.95\linewidth]{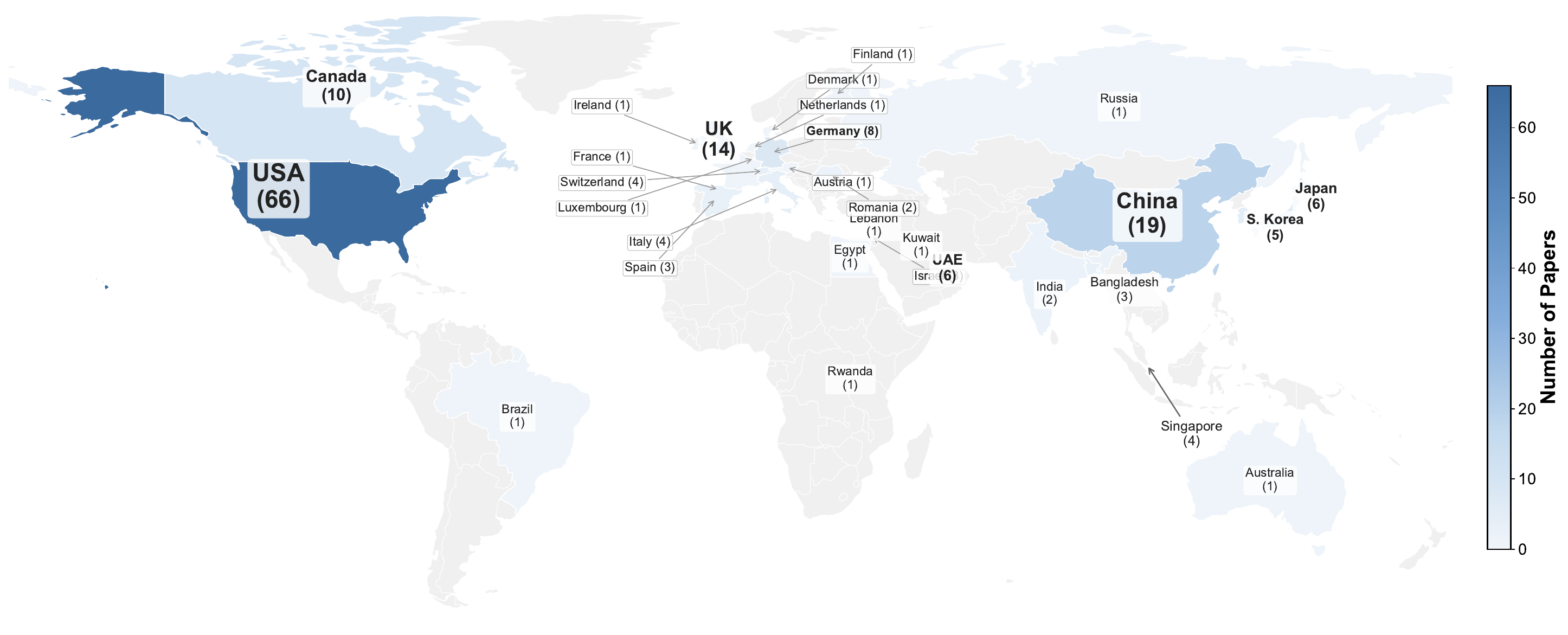}
\caption{\textbf{Geographic distribution of LLM-as-a-Judge studies in healthcare.} World map showing the number of publications by country, colored by paper count. The United States contributed the largest share (n=66, 48.9\%), followed by China (n=19), the United Kingdom (n=14), Canada (n=10), and Germany (n=8). Research contributions spanned 30 countries across six continents. Based on analysis of n=134 included studies.}
\label{fig:geo_map}
\end{figure*}

The geographic distribution of LLM-as-a-Judge research in healthcare reveals a pronounced concentration in the United States, which contributed to 66 of 134 studies (49.3\%), followed by China (n=19, 14.2\%), the United Kingdom (n=14, 10.4\%), Canada (n=10, 7.5\%), and Germany (n=8, 6.0\%) (Figure~\ref{fig:geo_map}). These five countries accounted for over 86\% of all publications. Research contributions spanned 30 countries across six continents, with notable representation from Asia-Pacific nations including Japan (n=6), the UAE (n=6), South Korea (n=5), and Singapore (n=4), reflecting the global interest in applying LLMs for clinical evaluation. European contributions extended beyond Germany to include Switzerland (n=4), Italy (n=4), Spain (n=3), and several other nations, indicating broad adoption across diverse healthcare systems.

\begin{figure*}[ht]
\centering
\includegraphics[width=1.0\linewidth]{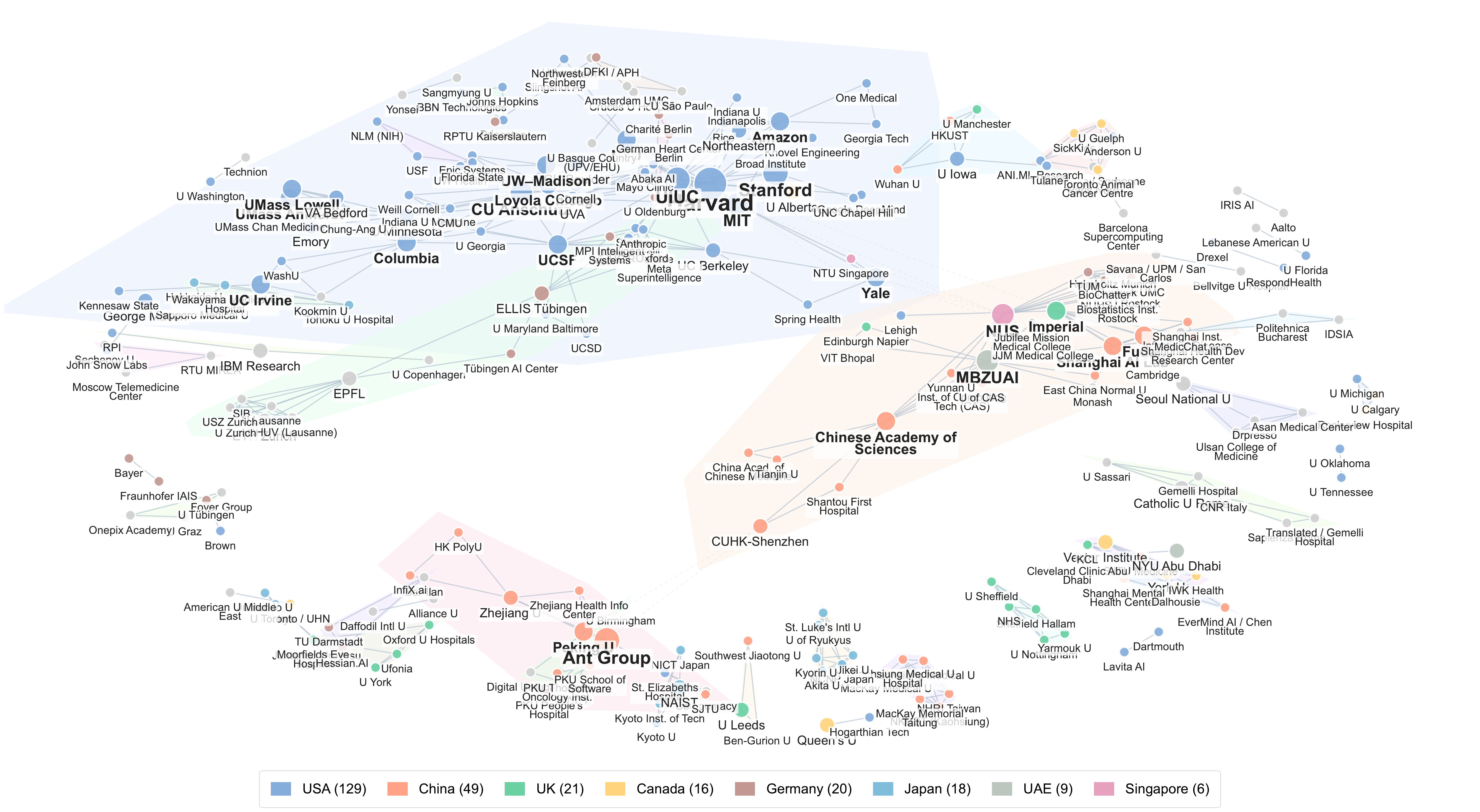}
\caption{\textbf{Institutional collaboration network for LLM-as-a-Judge research in healthcare.} Nodes represent institutions, sized by the number of contributing studies and colored by country/region. Edges indicate co-authorship on at least one study, with solid lines for intra-community collaborations and dashed lines for inter-community collaborations. Shaded regions denote communities identified by Louvain modularity optimization. Harvard University (n=8), Stanford University (n=5), UIUC (n=5), and Ant Group (n=5) were the most prolific contributors. Based on 241 institutions with at least one collaboration across 134 studies.}
\label{fig:collaboration}
\end{figure*}

The institutional collaboration network illustrates the multi-institutional nature of this research area (Figure~\ref{fig:collaboration}). Among 277 unique institutions represented across the included studies, Harvard University was the most prolific contributor (n=8 studies), followed by Stanford University, the University of Illinois Urbana-Champaign, and Ant Group (n=5 each), and MIT, the University of Colorado Anschutz, the National University of Singapore, and Mohamed bin Zayed University of Artificial Intelligence (n=4 each). Community detection analysis revealed distinct clusters of collaborating institutions, with a large US-centric cluster centered around Harvard, Stanford, and MIT, an Asia-Pacific cluster anchored by MBZUAI, NUS, and the Chinese Academy of Sciences, and a Chinese technology cluster led by Ant Group and Peking University, among others.

International collaboration was observed in 33 of 134 studies, with the most frequent cross-border partnerships occurring between the USA and Canada (n=4), China and the UK (n=4), and the USA and the UK (n=3). The predominance of US-based institutions and the relatively moderate rate of international co-authorship suggest opportunities for expanding cross-national collaboration, particularly between Western and Asian research groups, to address the diverse clinical and linguistic contexts in which LLM-based evaluation systems are deployed.

\section{Temporal Trends of Judge Models}
\label{app:model_temporal}

\begin{figure*}[ht]
\centering
\includegraphics[width=0.75\linewidth]{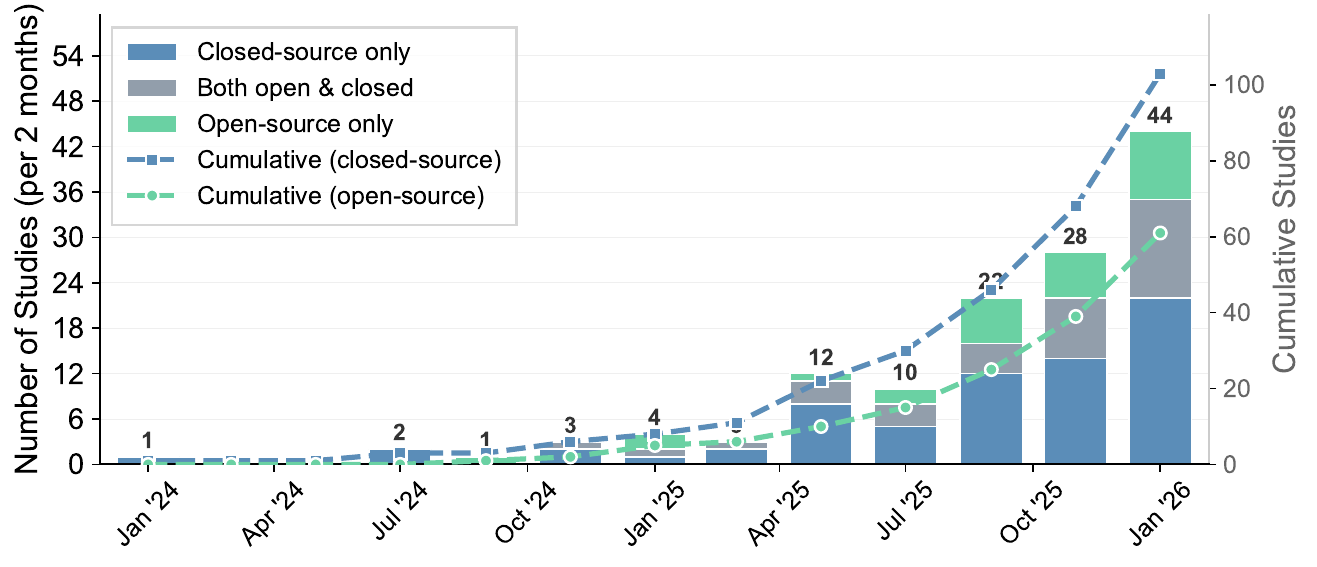}
\caption{\textbf{Temporal trends in open-source vs.\ closed-source LLM judge adoption.} Stacked bar chart showing the number of studies per bimonthly period classified by model type: closed-source only (blue), open-source only (green), and both (gray). Dashed lines show cumulative counts of studies using any closed-source model (blue) and any open-source model (green). Studies using both types are counted in both cumulative lines. N=130 studies with identifiable model types.}
\label{fig:model_temporal}
\end{figure*}

Among 130 studies with identifiable model types (4 studies do not clarify the judge models), 71 (54.6\%) use closed-source models exclusively, 26 (20.0\%) use open-source models exclusively, and 33 (25.4\%) use both (Figure~\ref{fig:model_temporal}). Overall, closed-source models appear in 104 studies (80.0\%) and open-source models in 59 studies (45.4\%). Bimonthly trend analysis shows that closed-source models dominate throughout the study period, but open-source adoption accelerates markedly from mid-2025 onward. In the most recent period (January--February 2026, n=44), 22 studies use closed-source models only, 13 use both, and 9 use open-source models only, indicating that 40\% studies incorporate open-source LLM judges.

The growing adoption of open-source models is possibly driven by the release of capable open-weight models such as \textit{DeepSeek-R1}, \textit{LLaMA-3.3-70B}, \textit{Qwen3-32B}, and \textit{Gemma-3-27B}, which approach proprietary model performance on many evaluation tasks. The concurrent rise of the ``both'' category reflects a trend toward multi-model evaluation pipelines that combine proprietary and open-weight judges, balancing evaluation quality against cost, reproducibility, and data privacy constraints. Open-source models are particularly attractive in healthcare settings where protected health information cannot be transmitted to external APIs, enabling on-premises deployment of judge models while maintaining evaluation reliability through ensemble configurations with closed-source models.

\end{document}